\documentclass{aa}  

\usepackage{graphicx}
\graphicspath{{../plots2/plots_final/}{../spec_plots/Pages/}}
\usepackage{epsfig}
\usepackage{color}
\usepackage{txfonts}
\usepackage{lscape}
\usepackage{captcont}
\usepackage{lscape}
\usepackage{longtable}
\usepackage{caption}
\usepackage{pdfpages}

\newcommand{\am}[2]{$#1'\,\hspace{-1.7mm}.\hspace{.0mm}#2$}
\newcommand{\HI}{H\textsc{i}}

\newcommand{\MHI}{$M_{\rm HI}$}
\newcommand{\Mz}{$M_{\rm z}$}
\newcommand{\Mg}{$M_{\rm g}$}
\newcommand{\Mstar}{$M_{\star}$}

\newcommand{\Msun}{$M_\odot$}
\newcommand{\kms}{\mbox{km\,s$^{-1}$}}
\newcommand{\nan}{Nan\c{c}ay}
\def\approxlt{\lower.2em\hbox{$\buildrel < \over \sim$}}
\def\approxgt{\lower.2em\hbox{$\buildrel > \over \sim$}}

\newcommand{\FHI}{\mbox{$F_{\rm HI}$}}

\newcommand{\Lsun}{\mbox{$L_{\odot}$}}

\newcommand{\VHI}{\mbox{$V_{\rm HI}$}}
\newcommand{\Vopt}{\mbox{$V_{\rm opt}$}}

\newcommand{\Wfifty}{\mbox{$W_{\mathrm 50}$}}
\newcommand{\Wtwenty}{\mbox{$W_{\mathrm 20}$}}
\newcommand{\Lr}{\mbox{$L_{\rm r}$}}
\newcommand{\SFR}{{\em SFR}}
\newcommand{\sSFR}{{\em sSFR}}

\newcommand{\SNR}{{\em SNR}}

\newcommand{\lamda}{\lambda}

\begin{document}

\offprints{Z. Butcher}

\title{NIBLES – an H{\bf \Large I} census of stellar mass selected SDSS galaxies: \\
II. Arecibo follow-up H{\bf \Large I} observations }
\author{Z. Butcher\inst{1}
  \and
S. Schneider\inst{1}
  \and
W. van Driel\inst{2,3}
  \and
M. D. Lehnert\inst{4}
  \and  
R. Minchin\inst{5}
}
  
\institute{University of Massachusetts, Astronomy Program, 536 LGRC, Amherst, MA 01003, U.S.A. 
\email{zbutcher@astro.umass.edu}
\and
  GEPI, Observatoire de Paris, CNRS, Universit\'e Paris Diderot, 5 place Jules Janssen, 92190 Meudon, France  
  \and
Station de Radioastronomie de \nan, Observatoire de Paris, CNRS/INSU USR 704, Universit\'e d'Orl\'eans OSUC, route de Souesmes, 18330 \nan, France 
	\and    
Institut d'Astrophysique de Paris, UMR 7095, CNRS Universit\'e Pierre et Marie Curie, 98 bis boulevard Arago, 75014 Paris, France  
  \and 
Arecibo Observatory, National Astronomy and Ionosphere Center, Arecibo, PR 00612, USA  
}

\abstract{
We obtained Arecibo \HI\ line follow-up observations of 154 of the 2600 galaxies in the \nan\ Interstellar Baryons Legacy Extragalactic Survey (NIBLES) sample.  These observations are on average four times more sensitive than the original observations at the \nan\ Radio Telescope.  The main goal of this survey is to characterize the underlying \HI\ properties of the NIBLES galaxies which were undetected or marginally detected at \nan.  Of the \nan\ non-detections, 85\% were either clearly or marginally detected at Arecibo, while 89\% of the \nan\ marginal detections were clearly detected.  Based on the statistics of the detections relative to $g$-$i$ color and $r$-band luminosity (\Lr) distribution among our Arecibo observations, we anticipate $\sim$60\% of our 867 \nan\ non-detections and marginal detections could be detected at the sensitivity of our Arecibo observations.  Follow-up observations of our low luminosity (\Lr\ < 10$^{8.5}$ \Lsun) blue sources indicate that they have, on average, more concentrated stellar mass distributions than the \nan\ detections in the same luminosity range, suggesting we may be probing galaxies with intrinsically different properties.  These follow-up observations enable us to probe \HI\ mass fractions, log(\MHI/\Mstar)  0.5 dex and 1 dex lower, on average, than the NIBLES and ALFALFA surveys respectively.}

   
\authorrunning{Z. Butcher et al.}
\titlerunning{NIBLES -- Arecibo follow-up \HI\ observations }
\maketitle

\section{Introduction}
\label{introduction}

The optical luminosity function (LF) and the \HI\ mass function (HIMF) are two of the most important and fundamental tracers of the volume density distribution of galaxies in the universe.  They yield clues to both the baryonic and dark matter content of galaxies, as well as their evolutionary histories.  Consequently, there are many applications for which the LF and HIMF can be used, for example, as constraints in galaxy formation models \cite[see, e.g.,][]{benson2003, lu2014}.

Many studies have attempted to constrain both of these functions over the years.  Since the LF was first fitted to an analytic form by \cite{schechter1976}, many subsequent studies have attempted to analyze its various properties and constrain its parameters (see, e.g., \citealt{felten1985, efstathiou1988, loveday1992, loveday2015,  blanton2001, blanton2003, dorta09, mcnaught2014}).  The HIMF, having the same functional form as the LF, has also been analyzed in detail, although to a somewhat lesser extent \cite[see, e.g.,][]{zwaan97, zwaan03, kilborn99, kovac2005, springob2005a, martin2010, hoppmann2015}.  

To date, both these functions have been treated separately in their analyses.  One of the main goals of the \nan\ Interstellar Baryons Legacy Extragalactic Survey (NIBLES) is to study the inter-relation between these two fundamental population tracers.  More specifically, we want to analyze the HIMF and other galaxy properties as a function of optical luminosity.  To achieve this goal, we carried out a 21cm \HI\ line survey at the 100m class \nan\ Radio Telescope (NRT). The final observed sample consists of 2600 galaxies selected from the Sloan Digital Sky Survey (SDSS; \citealp[see e.g.,][]{york00}) with radial velocities 900$<$cz$<$12,000 \kms.  The galaxies were selected to be distributed evenly over their entire range of absolute $z$-band magnitudes ($\sim-13.5$ to $-24$), which was used as a proxy for total stellar mass --- see \citet[Paper I]{vandriel2016} for further details. 

The NIBLES galaxy selection criteria are: 
\begin{enumerate}
\item{Must have both SDSS magnitudes and optical spectrum;}
\item {Must lie within the local volume (900$<$cz$<$12,000 \kms);}
\item {Uniform sampling of each 0.5 magnitude wide bin in absolute $z$-band magnitude, \Mz};
\item {Preferentially observe nearby objects;} 
\item {No a priori selection on color.}
\end{enumerate}

NIBLES, with its relatively uniform selection of galaxies that are based on total stellar mass, is aimed to complement other recent and/or ongoing large \HI\ surveys in the local volume, in particular, blind surveys such as ALFALFA \citep[e.g.,][]{haynes2011}.  One main advantage of NIBLES over blind \HI\ surveys is our increased on-source integration time, which enables us to reduce the $rms$ noise of the observations.  Each NIBLES source was initially observed at \nan\ for about 40 minutes of telescope time, resulting in a mean $rms$ noise of $\sim$3 mJy at $18$ \kms\ resolution.  In the case of weak or non-detections, observations were repeated (as time allowed) resulting in a target $rms$ noise between 1.5 and 1.8 mJy for the majority of our undetected sample.  However, there were a number of sources where follow-up time was unavailable to achieve the desired $rms$, which resulted in a mean of 2.3 mJy for the remainder of the undetected sample, yielding a bimodal $rms$ noise distribution (see Paper I). 

Of the 2600 NIBLES galaxies, 1733 (67\%) were clearly detected, 174 (7\%) marginally detected, and 693 (27\%) were not detected.  To adequately quantify our \HI\ distribution across the optical LF, we need to gain a statistical understanding of the underlying \HI\ distribution of sources which were undetected at \nan.  We therefore carried out pointed observations of 90 undetected or marginally detected galaxies at the 305m Arecibo radio telescope, which gives us a noise level reduction by about a factor of four.  Additionally, we had a number of sources suffering from observational problems at \nan\ which we re-observed at Arecibo, and during periods of time when primary target sources were unavailable, we observed detected NIBLES sources to compare flux calibrations at the two observatories.  In total, we observed 154 galaxies from the NIBLES sample (see Sect. \ref{sample selection} and Paper I for details).  

Here we present the results from these follow-up observations along with a brief synopsis of the differences in the data between the \nan\ and Arecibo samples.  The main purpose of this paper is data presentation.  Further analysis will be carried out in subsequent papers.  In Sect. \ref{sample selection} we describe the selection of the observed sample of galaxies and in Sect. \ref{observations}, the observations and data reduction. The results are presented in Sect. \ref{results} and discussed in Sect. \ref{discussion}.  An analysis of this data regarding the impact on our \HI\ distribution as a function of optical luminosity will be presented in Paper III (Butcher et al., in prep.).  All source numbers presented in this paper refer to the NIBLES source number, which can be cross-referenced with other common source names in the tables presented here and in Paper I.

\section{Sample selection} 
\label{sample selection}

Our total sample of Arecibo follow-up galaxies consists of 154 sources.  Of these, 90 are classified as either non-detections or marginal detections at \nan, with the remaining 64 consisting of sources initially suffering from observational problems such as OFF beam detections or RFI and a handful of sources used for flux comparison between the two telescopes.  Of these 64, ten were excluded from the original NIBLES catalog because their \nan\ observations contained technical problems which we were not able to overcome (listed in Table \ref{tab:additional}).

Of the 90 \nan\ non-detected or marginally detected galaxies, 59 were selected based on color ($u$-$z$ $<$ 2) and radial velocity ($cz$ < 4000 \kms) for the specific reason that these blue, nearby galaxies would normally be expected to have \HI\ and yet were undetected at \nan.

\section{Observations and data reduction} 
\label{observations}

The Arecibo radio telescope uses a 305m diameter spherical primary mirror and covers a declination range of $-1^{\circ} < \delta < 38^{\circ}$ with pointing accuracy of about 5$''$.  We used the L-band wideband receiver (L-wide) with the Wideband Arecibo Pulsar Processor (WAPP) correlator backend using two polarizations with a bandpass of 50 MHz (approximately 10,600 \kms) across 4096 frequency channels corresponding to a channel separation of 2.6 \kms at z$\sim$0.  The L-wide receiver has a half power beam width of approximately \am{3}{5} and yields an effective system temperature typically between 28 and 32 K.  Data were taken in standard 5/5 minute integration ON/OFF position switching mode.  All galaxies were observed for a minimum of one 5/5 minute cycle, and some of the blue low-luminosity galaxies were observed longer depending on telescope time and signal strength after the first observation. 

Throughout this paper, all radial velocities given are heliocentric,
and all \HI-line related parameters are according to the
conventional optical definition ($V$ = c($\lamda$ -- $\lamda_0$)/$\lamda_0$).

Observations were carried out in two sessions, between December 2008 and October 2009 and between March and September 2012 for a total of 59 hours.  

Data were reduced using a combination of Phil Perillat's IDL routines and Robert Minchin's CORMEASURE routine from the Arecibo Observatory.  All \HI\ spectra were Hanning smoothed to a median velocity resolution of 18.7 \kms\ to match the 18 \kms\ resolution of the NRT spectra as closely as possible.  All \HI\ spectra shown here have a heliocentric, optical ($cz$) radial velocity scale.

Two of our sources, 1260 and 2434 (i.e., \object{PGC 4546173} and \object{CGCG 427-032}), suffered from a baseline ripple with a wavelength corresponding to approximately 210 \kms\ which we were able to remove via a Fourier transform, see Appendix \ref{app:derip} for details.

\section{Results}  
\label{results}

The Arecibo observations enabled us to probe our sample about four times deeper on average than at \nan.  The mean $rms$ noise of the \nan\ undetected sample is 2.33 mJy whereas the mean $rms$ noise of our Arecibo observations is 0.57 mJy, both at 18 \kms\ resolution.

As with our \nan\ data, we divided the sources into detected, marginally detected, and non-detected categories.  This was accomplished through visual inspection of each \HI\ spectrum by three independent adjudicators (ZB, WvD, SES).  Disagreements were discussed until a consensus was reached.  The galaxies in the marginal detection category have \HI\ line spectra 
with a peak signal-to-noise ratio less than four, but coinciding with the SDSS optical velocity.  The galaxies in this category would most likely be missed in a survey of objects with previously unknown velocity.  However, the low probability of a strong noise peak coinciding with the SDSS optical velocity lends greater credibility to the likelihood that these peaks represent real signals.
 
There is generally very good agreement ($<$ 3 \kms\ on average) between the DR9 heliocentric velocities and our \HI\ velocities, with the exception of three cases.  These three outliers, sources 1631, 2434, and 2606 (i.e., \object{NGC 4290}, \object{CGCG 427-032}, and \object{NGC 3772} respectively) all have velocity discrepancies larger than 50 \kms.  Source 1631 has the SDSS spectral fiber positioned out in its disk, blueshifting the overall redshift measurement.  Source 2434 is confused with a secondary source in the Arecibo beam, but source 2606 suffers no obvious signs of confusion or spectral fiber offsets and the optical spectrum does not appear noisy (see Fig. \ref{AO_dets:5}).  However, source 2606 is also listed in \cite{de1991a} as having a heliocentric velocity of 3478$\pm$50 \kms\ which agrees with our \HI\ velocity of 3423$\pm$7 \kms.
 
The optical data listed are in general from the SDSS DR9 (see also Paper I) with the median total stellar masses and star-formation rates taken from the corresponding publicly available SDSS added-value MPA-JHU catalogs \citep{brinchmann04, kauffmann03b, salim07, tremonti04} where available.  In cases where estimates are not available, stellar masses, specific Star Formation Rates (\sSFR), and gas mass fractions (log(\MHI/\Mstar)) are marked as ``----''.  

Due to the limited redshift range of NIBLES, the difference in luminosity distances used for our \HI\ masses and the stellar mass estimates from MPA-JHU are less than a few percent in the most extreme cases.  This systematic difference is far less than the typical uncertainty in the stellar mass estimates themselves, which are on the order of 20\%.  

Listed throughout the tables are the following properties of the target galaxies:

\begin{itemize}
\item{RA \& Dec:} Right Ascension and Declination in J2000.0 coordinates, as used for the observations;
\item{Other name:} common catalog name, other than the SDSS;
\item{\Vopt:} heliocentric radial velocity ($cz$) measured in the optical (in \kms), from Paper I;
\item{log(\Mstar):} total median stellar mass estimates (in \Msun);
\item{log(\sSFR):} specific Star Formation Rate, or \SFR/\Mstar (in yr$^{-1}$);
\item{$g-z$:} $g$-$z$ integrated color of the galaxy using SDSS model magnitudes, corrected for Galactic extinction, following \citet{schlegel1998} (in mag);
\item{\Mg:} integrated absolute $g$-band magnitude, corrected for Galactic extinction following \citet{schlegel1998};
\item{$rms$:} $rms$ noise level values of the \HI\ spectra (in mJy);
\item{\VHI:} heliocentric radial velocity ($cz$) of the center of the \HI\ line profile (in \kms);
\item{\Wfifty, \Wtwenty:} velocity widths measured at 50\% and 20\% of the \HI\ profile peak level, 
respectively, uncorrected for galaxy inclination (in \kms);
\item{\FHI:} integrated \HI\ line flux (in Jy \kms);
\item{$SNR$:} peak signal-to-noise ratio, which we define as the peak flux density divided by the $rms$;
For non-detections, the $SNR$ listed is the maximum found in the expected velocity range of the \HI\ profile; 
\item{$S/N$:} signal-to-noise ratio determined taking into account the line width, following the
ALFALFA \HI\ survey formulation from \cite{saintonge07}: 
S/N = 1000(\FHI/\Wfifty)$\cdot$(\Wfifty/2$\cdot$R)$^{0.5}$)/$rms$, where $R$ is the velocity resolution, 18.7 \kms on average;
\item{log(\MHI):} total \HI\ mass (in \Msun), where \MHI\ = 2.36$\times$$10^5$$\cdot$D$^2$$\cdot$\FHI, where $D$ = $V$/70 is the galaxy's distance (in Mpc). In the cases of non-detections, 3$\sigma$ upper limits are listed for a flat-topped profile with a width depending on the target’s $r$-band luminosity, \Lr, according to the upper envelope in the \Wtwenty\ - \Lr\ relationship of our \nan\ clear, non-confused detections (see Paper I);
\item{log(\MHI/\Mstar):} ratio of the total \HI\ and stellar masses.
\end{itemize}

Estimated uncertainties are given after the values in the tables. Uncertainties in the central \HI\ line velocity, \VHI, and in the integrated \HI\ line flux, \FHI, were determined following \citet{schneider86, schneider90} as, respectively 
\begin{equation} 
\sigma_{V_{HI}} = 1.5(W_{20}-W_{50})S\!NR^{-1}\, (\kms) 
\end{equation} 
\begin{equation} 
\sigma_{F_{HI}} = 2(1.2W_{20}R)^{0.5}rms\, (\kms) 
\end{equation} 
where $R$ is the instrumental resolution, 18 \kms, \SNR\ is the peak signal-to-noise ratio of a spectrum and $rms$ is the root mean square noise level (in Jy).  Following Schneider et al., the uncertainty in the \Wfifty\ and \Wtwenty\ line widths is expected to be 2 and 3.1 times the uncertainty in \VHI, respectively.

Table \ref{tab:AOdetNRTnd} lists all 72 sources detected at Arecibo which were either undetected (55 sources) or marginally detected (17 sources) at \nan.  Table \ref{tab:AOmarNRTnd} lists the five Arecibo marginal detections which were undetected at \nan, and Table \ref{tab:AO_ND} lists the 15 sources not detected at Arecibo, including 11 \nan\ non-detections, two \nan\ marginal detections and two others flagged as NRT confused detections which were not detected at Arecibo due to its smaller beam size.

In Table \ref{tab:bothDETs} we compare line flux parameters of galaxies we detected at both Arecibo and \nan\ and in Table \ref{tab:additional} we list the ten sources detected at Arecibo which were not included in our final \nan\ sample due to data problems (see Paper I). 

The following types of cases have been flagged in the tables following the naming conventions in Paper I:

\begin{itemize}
\item C (3 cases): \HI\ detection of the target galaxy confused by another galaxy within the Arecibo telescope beam;
\item C3 (1 case): \HI\ detection of the target galaxy confused by another galaxy in the Arecibo telescope beam, but the secondary source likely contributes a minor amount of flux to the total observed flux;
\item K (8 cases): \nan\ \HI\ detection either clearly or possibly confused by another galaxy within the NRT beam;
\item D (2 cases): baseline ripple removed from \HI\ spectrum (see Sect. 3);
\item M (19 cases): original \nan\ detection classification changed to marginally detected;
\item R (2 cases): sources possibly resolved by the Arecibo beam.  These sources have an SDSS optical image with a diameter about the same size as the Arecibo beam which is expected to lead to an underestimate of their total \HI\ flux.  
\end{itemize}

Color SDSS images alongside the \HI\ line spectra of our Arecibo detections are shown in Fig. \ref{AO_dets:1}, marginal detections are presented in Fig. \ref{AO_mars}, and non-detections in Fig. \ref{AO_ND}.  

Due to the updated detection category classifications and reprocessing of the \nan\ data, our Arecibo follow-up sample now consists of 54 galaxies detected at \nan\ (of which 52 were detected at Arecibo and the remaining two are confused detections at \nan\ where the signal is from a secondary source) as well as 90 galaxies which are either undetected (71) or marginally detected (19) at \nan\ and ten galaxies for which we have no useable \nan\ data.  These ten have been added to the NIBLES catalog and assigned succeeding source numbers, see Table \ref{tab:additional}.

\subsection{Flux comparison between Arecibo and \nan}
\label{AONRTdets}

For our ten calibration sources which have a \nan\ peak signal-to-noise ratio greater than 8, we compared the \HI\ line flux ratios at both telescopes using a weighted mean.  Each source was weighted by the inverse square of its flux uncertainty to more heavily weight the sources with lower errors.  The resulting Arecibo/NRT flux ratio is 1.19$\pm$0.08 where the uncertainty is given by the standard deviation of the weighted mean.  This offset is consistent with the flux offset discussed in Paper I.  Detailed analysis of this flux scale difference is beyond the scope of this paper (see Paper I for details).

\section{Discussion} 
\label{discussion}
\subsection{Arecibo detection rate of \nan\ non-detections and marginal detections}
\label{AOdetectionrate}

In this section and the remainder of this paper we focus on the Arecibo follow-up sample consisting of the 90 galaxies observed at Arecibo which were either undetected or marginally detected at \nan. The \nan\ marginal detections are included with the \nan\ non-detections for the analyses presented below since the marginal detection category had not yet been implemented at the time of this our Arecibo survey.

\begin{figure}
\centering
\includegraphics[width=9cm]{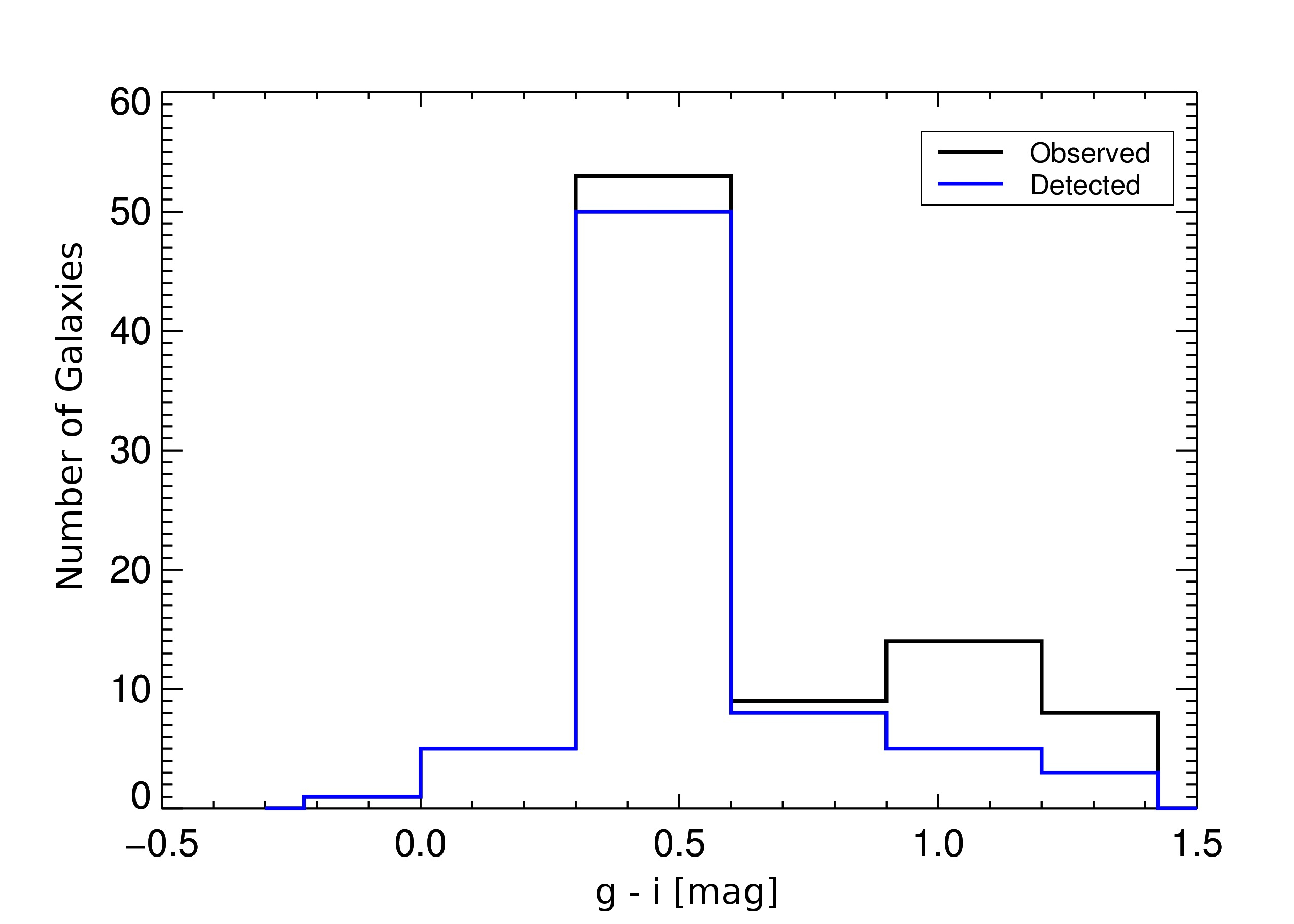}
\caption{\small Number of galaxies as a function of $g$-$i$ color, corrected for Galactic extinction following \citet{schlegel1998}.  The black line is the total number of observed galaxies in each 0.25 mag wide bin and the blue line the total number of detected galaxies.  The peak around 0.5 mag is due to our specific selection of blue objects (see Sect. 2).}
\label{detnum}
\end{figure}

Of our 90 \nan\ non-detections or marginal detections, 72 had sufficient \HI\ line flux to be clearly detected at Arecibo, resulting in a high overall detection rate of 80\%  (greater than 85\% counting Arecibo marginal detections).  One likely reason for this high detection rate is that most of our sample consisted of predominately blue, nearby star-forming galaxies.  This color bias is obvious in Fig. \ref{detnum},
which shows the number of galaxies observed and detected as a function of $g$-$i$ color.  All of the galaxies with $g$-$i$ colors bluer than 0.3 were detected, whereas only one third of the reddest galaxies were detected.

\begin{figure}
\centering
\includegraphics[width=9cm]{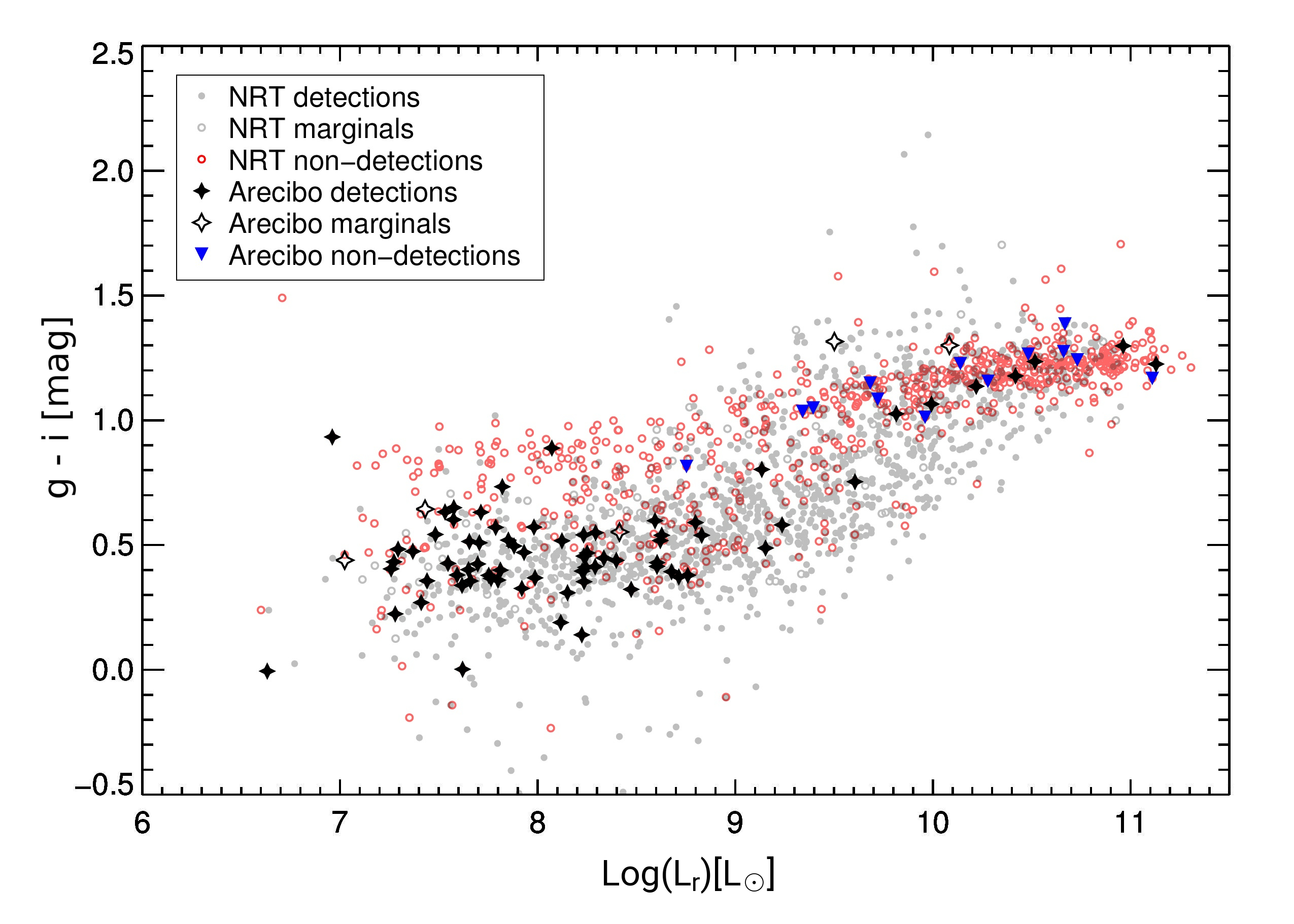}
\caption{\small Integrated $g$-$i$ color, in mag, as function of absolute $r$-band luminosity, log(\Lr) in \Lsun, both corrected for Galactic extinction following \cite{schlegel1998}.  \nan\ detections, marginals, and non-detections are represented by gray dots, open gray circles, and open red circles respectively.  Arecibo detections, marginals, and non-detections are respectively represented by black solid stars, open stars, and blue downward triangles.  The low luminosity end (log(\Lr) $<$ 9) shows a color dichotomy with \HI\ detections and non-detections clustering around $g$-$i$ $\sim$0.4 and $g$-$i$ $\sim$0.8 respectively, showing a clear distinction between the blue cloud and red sequence galaxies.  Above luminosities of log(\Lr) $\sim$9.5, the blue cloud disappears while the red sequence galaxies shift to redder colors, $g$-$i$ $\sim$1.2.}

\label{g_i_Lr}
\end{figure}

A color-magnitude diagram of integrated $g$-$i$ colors as a function of $r$-band luminosity is shown in Fig. \ref{g_i_Lr} for our detections, non-detections, and marginals of both the Arecibo and the \nan\ samples.  The dichotomy in both color and luminosity between \HI\ detections and non-detections is obvious in this plot.  At the low luminosity end (log(\Lr) < 9), the \nan\ data show a color dichotomy with the detections clustering around $g$-$i$ $\sim$0.4 (commonly referred to as the ``blue cloud'') and non-detections clustering around $g$-$i$ $\sim$0.8 (commonly referred to as the ``red sequence'').  This dichotomy has not previously been seen in larger surveys such as \cite{gavazzi2010, huang2012}, primarily due to the smaller dynamic range probed by the former and by the exclusion of lower luminosity red galaxies in the latter.  However, due to the NIBLES selection criteria and our $\sim$1 dex larger dynamic range, our sample contains a clear separation by color at low luminosities.

The sole Arecibo non-detection in this low-luminosity range (inverted blue triangle in Fig. \ref{g_i_Lr}) has a relatively red color, but several other low-luminosity red sources were detected, and two of the bluer low-luminosity galaxies were only marginally detected at Arecibo.  Further follow-up of these red, low luminosity sources is needed to establish the \HI\ mass properties of this low-luminosity red population.

The blue cloud largely disappears at luminosities higher than log(\Lr) = 9.5, where the majority of sources cluster around $g$-$i$ color of $\sim$1.2.  However, aside from the fact that these galaxies have relatively more non-detections compared to the low luminosity galaxies, the mixture of detections with non-detections suggests that a not insignificant fraction of the red sequence galaxies may contain detectable levels of \HI.  However, deeper follow-up observations of red sequence galaxies will be needed to answer this question.

\begin{figure}
\centering
\includegraphics[width=9cm]{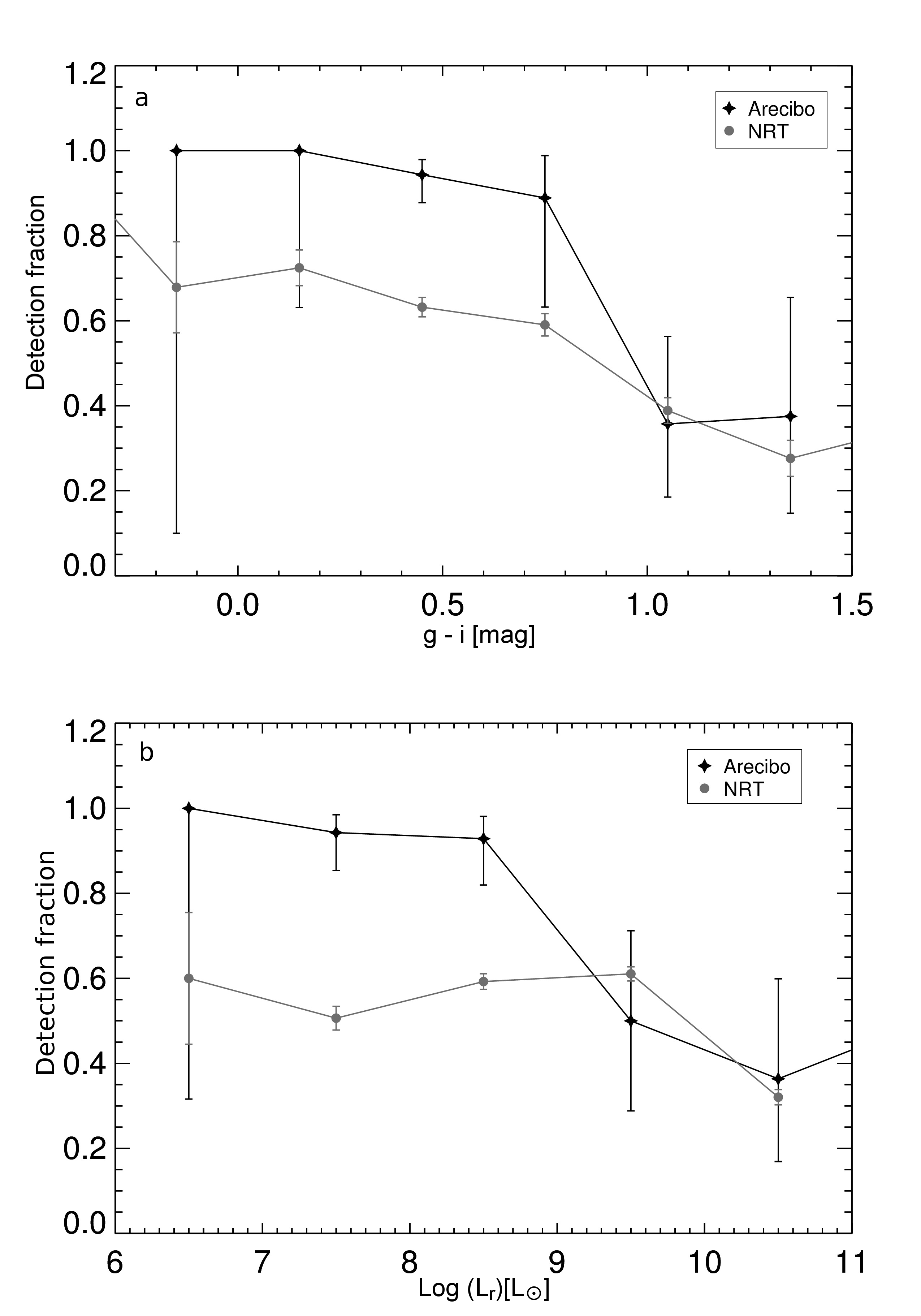}
\caption{\small {\bf a.} Detection fraction for the NIBLES \nan\ sample (gray) and the Arecibo follow-up sample (black) as a function of $g$-$i$ color.  {\bf b.} Detection fractions as a function of $r$-band luminosity.  Error bars for the \nan\ data are the standard deviation of the binomial distribution and the error bars for the Arecibo data were calculated following \citet{gehrels1986} using the 90\% confidence intervals.  Bin sizes are 0.3 for {\bf a} and 1 for {\bf b}.}
\label{detfrac}
\end{figure}

We compare the detection fractions as a function of $g$-$i$ color and \Lr\ for our \nan\ and Arecibo samples in Fig. \ref{detfrac}, counting the marginal detections in with the non-detections.
The \nan\ data show a global decrease in detection fraction as a function of color while the Arecibo data show a rather sharp drop by about a factor of 2.5 above $g-i$ $\sim$0.8.  As a function of luminosity, the \nan\ detection percentage shows no decrease below log(\Lr) $\sim$10 while the Arecibo sample shows a two times lower detection rate above log(\Lr) $\sim$9.  The plotted uncertainties in the \nan\ data points are the standard deviation of the binomial distribution, given by

\begin{equation}
\sigma = \sqrt{\frac{P(1-P)}{n}}
\label{sdbinomial}
\end{equation}
where P is the probability of detection, given by $m/n$ where $m$ is the number of detections and $n$ is the total number in a particular bin.  The Arecibo data generally have a small number of sources per bin which in some cases makes the uncertainty difficult to quantify.  We therefore adopted the 90\% confidence limits from \citet{gehrels1986} for dealing with the small number statistics of the Arecibo sample.

\begin{figure}
\centering
\includegraphics[width=9cm]{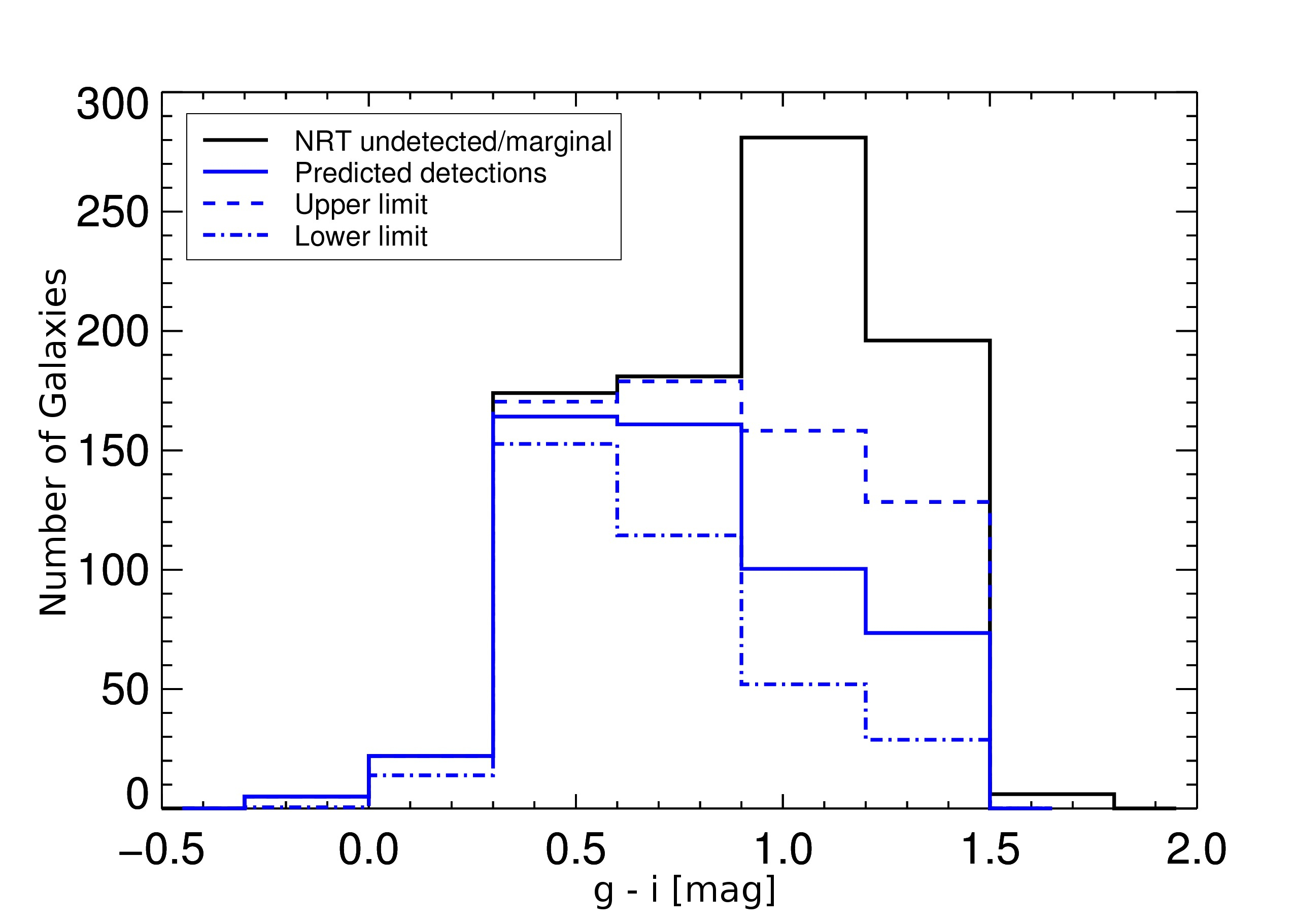}
\caption{\small Total number of galaxies either undetected or marginally detected at \nan\ that would be detectable with the same sensitivity as our Arecibo observations, as a function of $g$-$i$ color.  Black indicates \nan\ non-detections and marginals, whereas the solid blue line is the estimated number of galaxies that would be detectable based on our Arecibo sample detection rate.  The blue dashed and dash-dotted lines are based on, respectively, the upper and lower envelopes of the uncertainties in the detection rate, see Fig. \ref{detfrac}}

\label{detpredict}
\end{figure}

If we apply the Arecibo detection percentages to the \nan\ sample, we would expect (with observations of the same sensitivity as our Arecibo sample, i.e., four times lower noise than at \nan) to detect about 60\% (or $\sim$530) of the 867 \nan\ non-detections and marginal detections: based on color statistics the expected number is, $526\pm160$ and it is $546\pm158$ from the luminosity data (see Fig \ref{detpredict}). 
The uncertainty is calculated by adding in quadrature the fractional uncertainty in each $g$-$i$ or \Lr\ bin, respectively.

We have a total remaining sample of 442 \nan\ undetected or marginally detected galaxies within the Arecibo declination range, consisting mostly of redder low-luminosity (below \Lr\  $\sim$ $10^9$ \Lsun) sources as well as some higher luminosity red sequence galaxies.  Based on our current results, we estimate that comparable observations of the remaining 442 sources at Arecibo would require about 160 hours of telescope time: 60\% to be observed for our standard 5 minute on-source integrations and the other 40\% for 15 minutes each.  If detection statistics follow the same pattern as this sample, we would expect about 60\% of the remaining \nan\ undetected sample to be detected at Arecibo.  This would bring the over-all detection rate of the NIBLES sample within the Arecibo declination range to $\sim$89\%, and the global detection rate to  $\sim$77\%.  If we extrapolate these detection percentages to the entirety of the NIBLES sample, we would expect the NIBLES global detection rate to increase to $\sim$86\%.

Of the subset of our sample that was selected on low luminosity and blue color, we achieved a 100\% detection rate (including the Arecibo marginal detections).  We also managed to detect several red sequence galaxies with very low \HI\ stellar mass ratios, log(\MHI/\Mstar) $< -2$ (see Fig. \ref{MHI_Mstar}).  Two of these have stellar masses greater than $10^{11.5}$ \Msun, placing them in an area of \MHI/\Mstar--\Mstar parameter space not probed by the ALFALFA survey and as yet virtually unexplored.

\subsection{Physical properties of Arecibo-detected galaxies}
\label{propertiesAOdets}

\begin{figure}
\centering
\includegraphics[width=9cm]{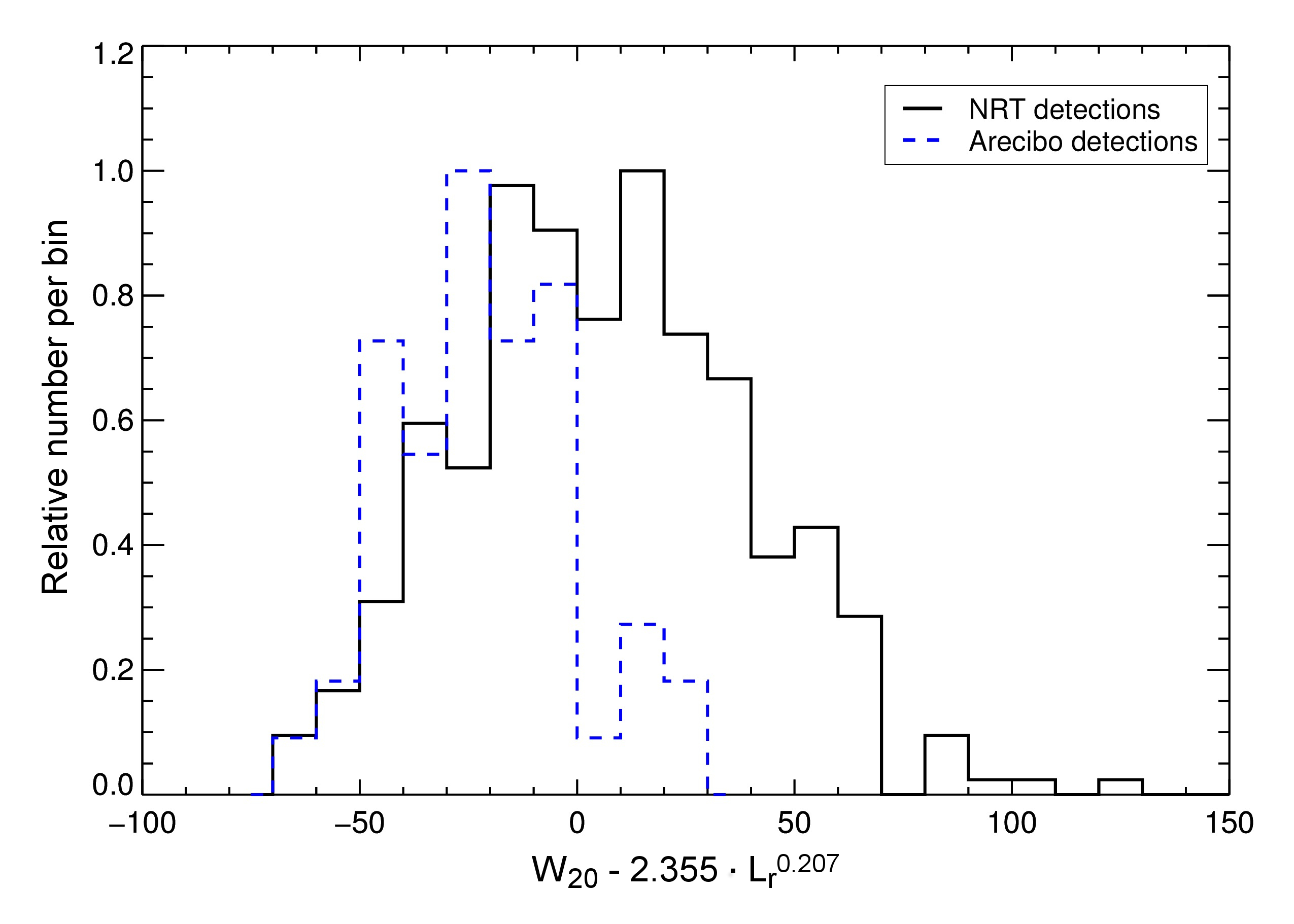}
\caption{\small Relative number of galaxies per bin for the \nan\ and Arecibo samples as a function of the fit to the \Wtwenty\ - \Lr\ relationship, i.e., \Wtwenty\ -- 2.355 $\cdot$ \Lr$^{0.207}$ with a bin size of 10.}
\label{W20_hist}
\end{figure}

In Paper I we used the \Wtwenty\ - \Lr\ relationship to estimate the maximum \HI\ line width a galaxy typically has for a given luminosity.  The least-squares fit to this relationship is log(\Wtwenty) = $0.4 + 0.2 \cdot \textrm{log(\Lr)}$.  
When comparing the Arecibo data to the \nan\ data, the Arecibo detections appear to have \Wtwenty\ values that are about 35\% narrower than the corresponding \nan\ detections of the same luminosity.  This is due in large part to our selection of low luminosity blue dwarf galaxies, which are predominately supported by velocity dispersion rather than rotation.  Consequently, the \HI\ line profiles of these galaxies are typically Gaussian shaped rather than displaying the commonly seen two-horned profiles.  Since these galaxies make up $\sim$70\% of our Arecibo detections, they are primarily responsible for this offset.  To illustrate this effect, we subtract the fit to the \Wtwenty\ - \Lr\ relationship from each source in the range log(\Lr) $\leq$ 8.5 and plot the resulting distributions for both the \nan\ and Arecibo data in Fig. \ref{W20_hist}.  The mean and standard deviations of the \nan\ and Arecibo distributions are 7$\pm$36 and --22$\pm$20 \kms\ respectively.  As is evident, the majority of the line widths for the Arecibo sample lie below the mean fit of the \nan\ sample and have a much narrower distribution.

To see if this difference in \HI\ line profile width corresponds to any differences in over-all stellar distribution, we examine the distributions of $r$-band half-light radius.  Here we define the half-light radius as the $R_{50}$ radius encompassing 50\% of the Petrosian flux, scaled for the Hubble-flow distance to the galaxy.  This radius gives a general sense of the stellar mass concentration within a galaxy under the assumption that the mass-to-light ratios of galaxies of the same luminosity are fairly consistent.  To test this assumption, we examined the distribution of $g$-$r$ colors between the low luminosity (log(\Lr) $\leq$ 8.5) Arecibo and NRT samples and found no significant difference (a Kolmogorov-Smirnov (K-S) test returns a 49.4\% probability that these distributions are not drawn from the same parent sample).  The similarities in color distribution between the two samples indicates that they do not have vastly differing stellar populations.  Therefore, differences in half-light radius should indicate differences in compactness. 

\begin{figure}
\centering
\includegraphics[width=9cm]{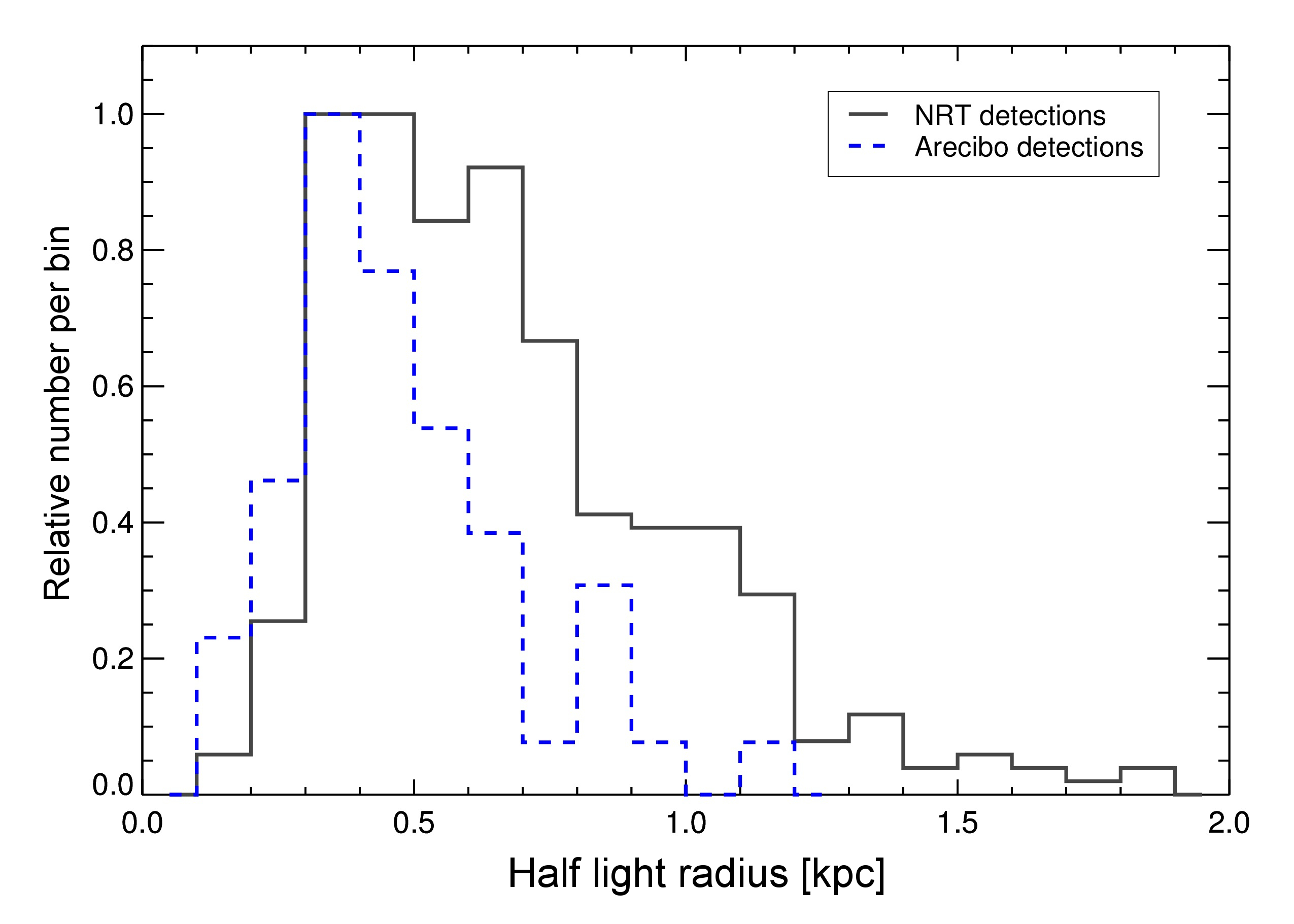}
\caption{\small Relative number of galaxies per bin of the \nan\ and Arecibo samples as a function of half-light radius, in kpc.  Bin size is 100 pc.}

\label{radius_hist}
\end{figure}

We plot the half-light radius distribution of the two samples in Fig. \ref{radius_hist}, which shows that the Arecibo data are systematically offset to lower radii than the \nan\ data.  The mean and standard deviations are 0.68$\pm$0.32 kpc for \nan\ and 0.47$\pm$0.22 kpc for Arecibo.  A K-S test returns a 94\% probability that the \nan\ and Arecibo samples are not drawn from the same parent distribution.

The smaller half-light radii of the Arecibo sample, combined with the fact that these galaxies have narrower \HI\ line widths, on average, may indicate that these galaxies are indeed more compact.  Since dwarf galaxies tend to have rotation curves that increase with radius, the narrow \HI\ line widths of these galaxies may be an indicator that the \HI\ is being confined to the central regions of the galaxy.  If this is the case, it may imply that these galaxies are in a different phase of evolution than their \nan\ detected counterparts.  However, further investigation is necessary.

\begin{figure}
\centering
\includegraphics[width=9cm]{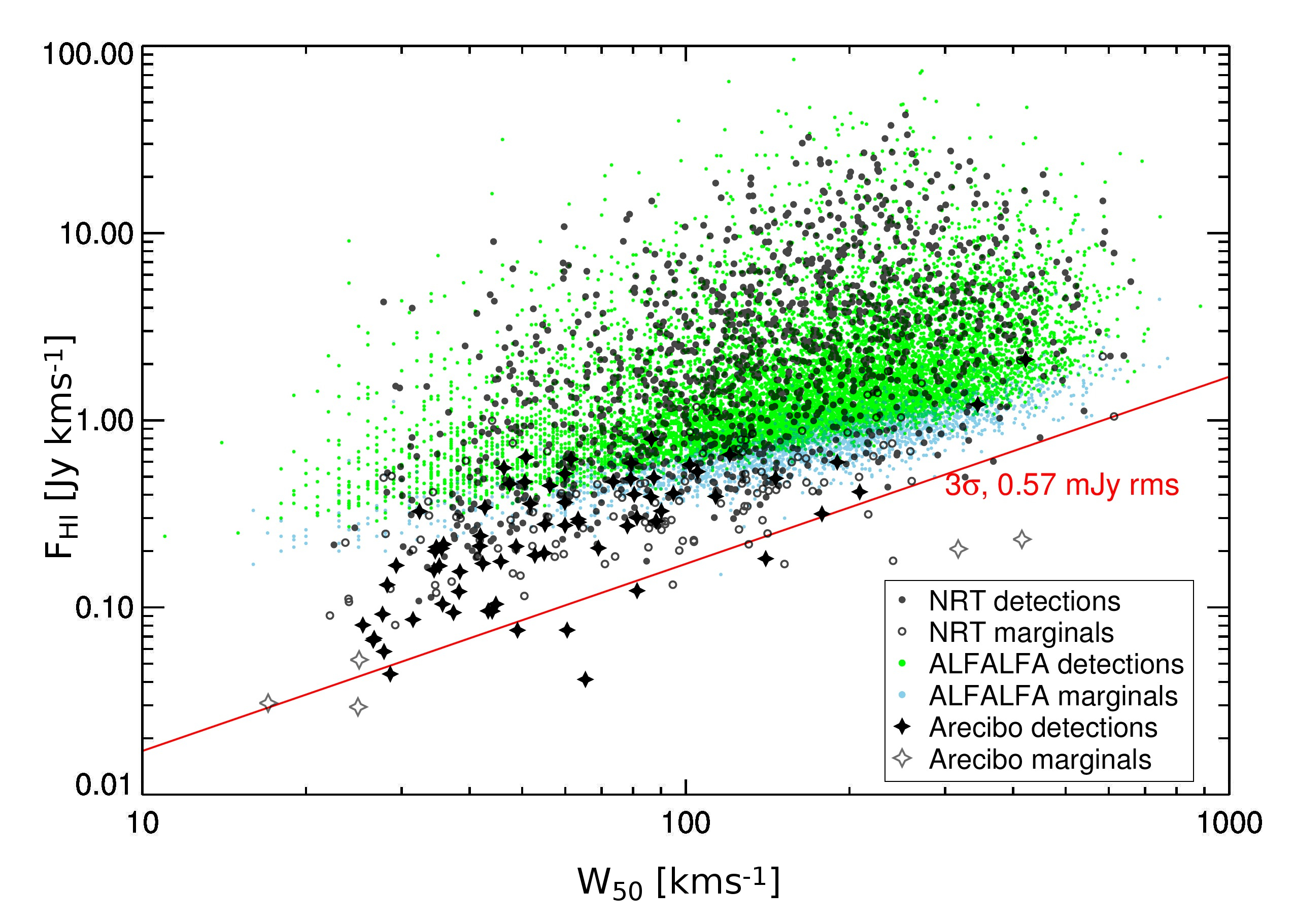}
\caption{\small Integrated \HI\ line flux \FHI\ (in Jy \kms) as a function of \Wfifty\ line width (in \kms). Shown are the NIBLES non-confused and non-resolved detections and marginals from both our Arecibo and \nan\ observations, together with the ALFALFA data from the $\alpha.40$ catalog \citep{haynes2011}, where green dots represent detections (their Category 1) and light blue dots represent marginals (Category 2). The red line indicates the integrated line flux from a 3$\sigma$ flat-topped profile with a 0.57 mJy $rms$ noise level.}

\label{FHI_W50}
\end{figure}

In Fig. \ref{FHI_W50} we show the integrated \HI\ line flux as a function of the \Wfifty\ line width for both our Arecibo and \nan\ non-confused and non-resolved detections and marginals, together with the ALFALFA detections and marginals from \cite{haynes2011} for comparison.  We also indicate the line flux for a flat-topped 3$\sigma$ detection with the mean Arecibo $rms$ noise level of 0.57 mJy, to show where weak detections are expected to lie.  

The ALFALFA data show an increasingly marked absence of weak sources with decreasing line width, beginning at about \Wfifty$\sim$100 \kms.  This is to due to both the detection threshold of ALFALFA (see \citealt{giovanelli05b}) and the fact that blind surveys have no a priori knowledge of source redshifts.  In the NIBLES observations, prior knowledge of the source redshift enables us to identify signals to a lower $SNR$.

\begin{figure}
\centering
\includegraphics[width=9cm]{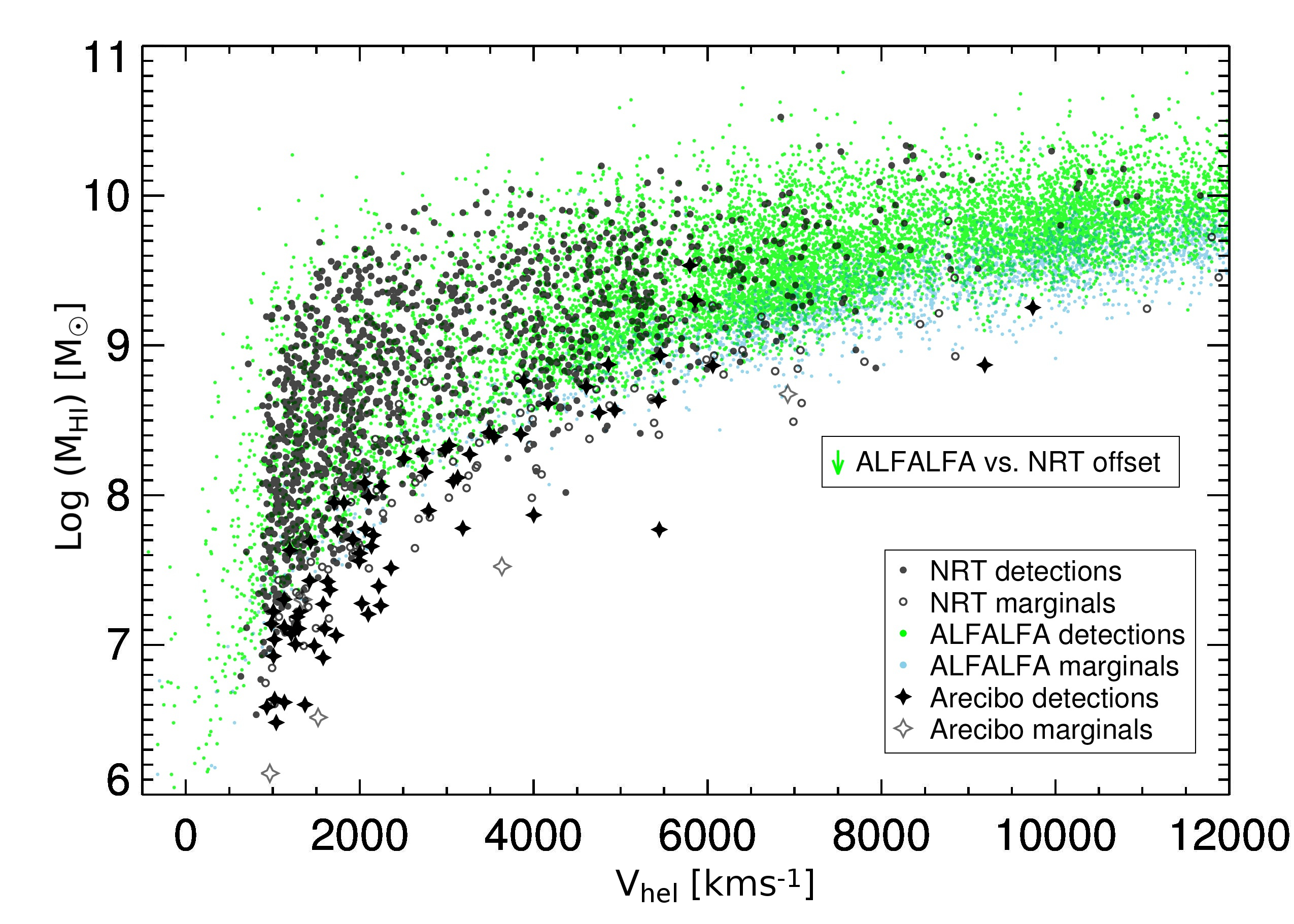}
\caption{\small Total \HI\ mass, \MHI\ (in \Msun), as a function of \HI\ radial velocity (in \kms). Shown are the NIBLES detections and marginals from both our Arecibo and \nan\ observations (black stars and dots respectively), together with data based on the ALFALFA $\alpha.40$ catalog \citep{haynes2011}, where green dots represent detections (their Category 1) and light blue dots represent marginals (Category 2). Excluded were all NIBLES detections which are definitely or probably confused by another galaxy within the telescope beam, as well as detections that are likely resolved (see Paper I). The \HI\ masses of the ALFALFA detections were calculated in the same way as for the NIBLES sources, using simply a distance of $D = V/H_0$, where the adopted Hubble constant is $H_0$ = 70 \kms\ Mpc$^{-1}$.  The green vertical arrow above the legend indicates the difference of 0.16 in log(\MHI) corresponding to the mean \HI\ flux scale difference between the $\alpha.40$ catalog and our NRT data.}

\label{MHI_VHI}
\end{figure}

Fig. \ref{MHI_VHI} shows \HI\ masses as a function of radial velocity for the same sources shown in Fig. \ref{FHI_W50}.  Additionally, we recalculated the \HI\ masses of the ALFALFA galaxies using pure Hubble-flow distances in order to maintain consistency with the NIBLES calculations.  The green vertical arrow represents the 0.16 dex average offset in \HI\ mass due to difference in flux scale between the $\alpha.40$ catalog and our NRT data discussed in Paper I. 

\begin{figure*}
\centering
\includegraphics[width=18cm]{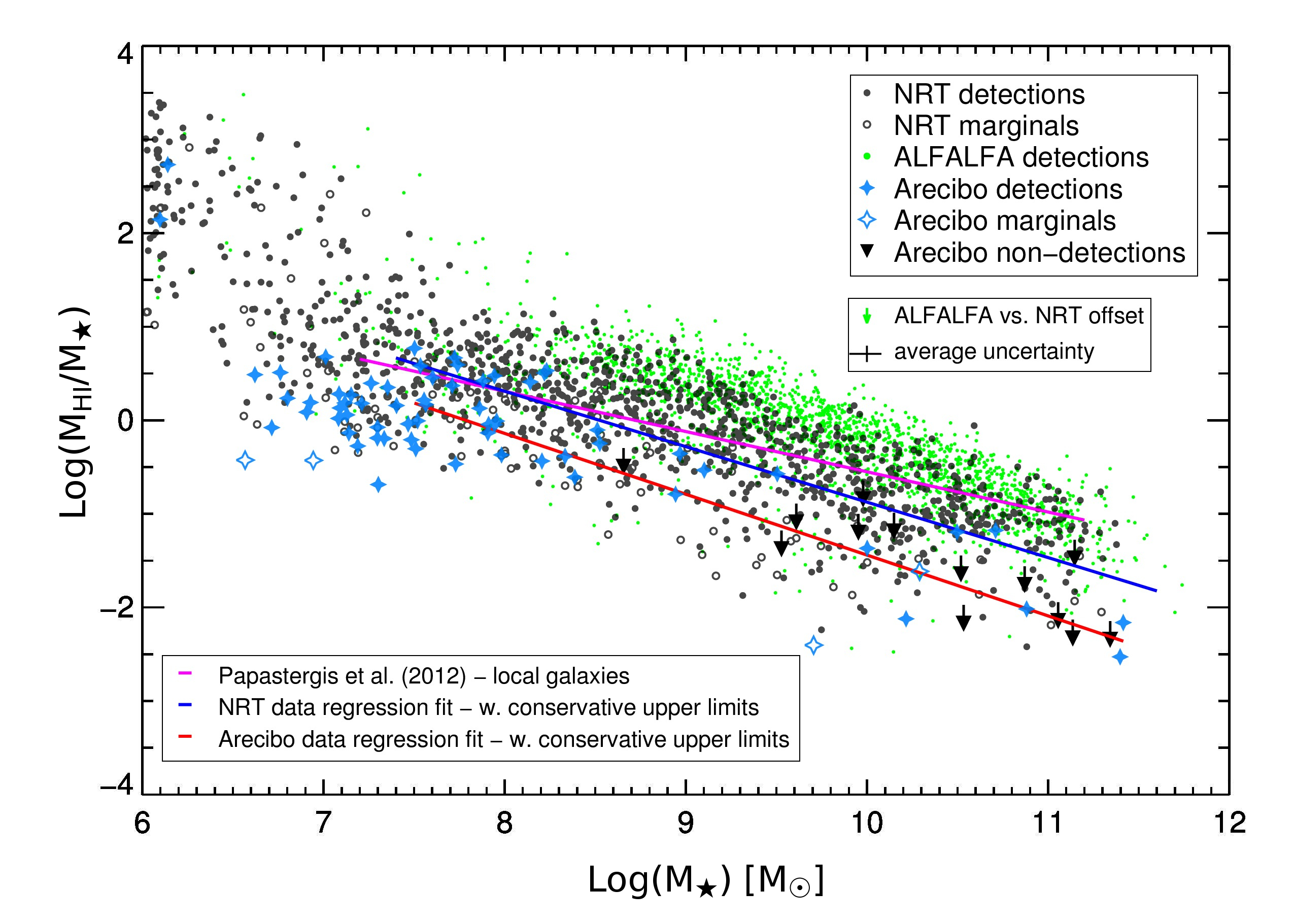}
\caption{\small \HI\ mass fraction (\MHI/\Mstar) as a function of total stellar mass, \Mstar (in \Msun).  NIBLES \nan\ detections and marginals are represented by gray dots and open circles respectively while ALFALFA detections are represented by green dots.  Our follow-up Arecibo detections and marginals are represented by solid and open light blue stars respectively while upper limits for non-detections are shown as downward arrows.  Stellar masses used were taken from the SDSS added-value MPA-JHU catalogs (see Sect. \ref{results}).  The green arrow below the legend represents the 0.16 dex average offset between the ALFALFA and NIBLES \HI\ masses due to flux scale differences and the black cross indicates the average uncertainty in the stellar masses and mass fractions (about 20\% in both cases).  The magenta line represents the fit to literature reference samples of \HI-detected galaxies from \citet{papastergis2012}, the blue and red lines represent the regression fits to the \nan\ and Arecibo sources, respectively, which include estimated upper limits to \HI\ masses of undetected galaxies (see Paper I for further details).}

\label{MHI_Mstar}
\end{figure*}

In Fig. \ref{MHI_Mstar} we compare the \HI\ mass fraction log(\MHI/\Mstar) as a function of stellar mass (log(\Mstar)), showing our \nan\ (NRT) detections and marginals, ALFALFA detections, and our Arecibo sample detections, marginals, and upper limits for the non-detections.  Our Arecibo upper limits were calculated following the same method used in Paper I, i.e., using a line width estimate from the upper envelope of the \Wtwenty\ - \Lr\ relationship.  For ALFALFA, we also took the stellar mass estimates from the DR9 added-value MPA-JHU catalogs for consistency with the NIBLES sample (see Paper I).

The distribution of log(\MHI/\Mstar) vs. log(\Mstar) data shows the same general trend as presented in Paper I and \cite{papastergis2012}, with the \HI-selected galaxy samples lying at systematically higher \HI\ mass fractions for a given stellar mass than the optically selected NIBLES galaxies.  We note that \cite{papastergis2012} used their own method for estimating the stellar masses of their sample (see \citealp{huang2012} for details).  As mentioned in Paper I, the ALFALFA fluxes are systematically higher than ours by a factor of 1.45 due to flux scale differences (see the 0.16 dex vertical green arrow in Fig. \ref{MHI_Mstar}). However, even taking this offset into account, the \HI-selected sample is has systematically higher \MHI/\Mstar ratios for a given \Mstar than NIBLES.  

To assess the parameter space probed by our Arecibo sample, we also compare the literature fit made in \cite{papastergis2012} of the four reference samples with \HI\ detections they used to evaluate the gas-to-stellar mass ratios of local galaxies (magenta), with fits to our data.  We used the Buckley-James method of linear regression, taking into account the \HI\ non-detections, from the STSDAS statistics package\footnote[1]{http://stsdas.stsci.edu/cgi-bin/gethelp.cgi?statistics} to fit the \nan\ (blue line) and Arecibo (red line) samples (see also Paper I).  Since the log(\MHI/\Mstar) vs. log(\Mstar) relationship becomes non-linear below log(\Mstar) $\sim$10$^{7.5}$, we only include masses above this cut-off in our fits.  Average uncertainties for each Arecibo source are represented by the cross below the legend.  We estimate a mean 1$\sigma$ stellar mass uncertainty of about 20\%, based on \citet{kauffmann03b}, and a typical uncertainty of about 20\% for the NIBLES Arecibo gas mass fractions.  Our typical gas mass fraction uncertainty for the entire NIBLES sample is about 27\% (see Paper I for details). 

As mentioned in Paper I, there is an increasing discrepancy between the \citet{papastergis2012} reference sample \HI\ mass fraction (magenta) and the \nan\ mass fraction (blue) as a function of increasing stellar mass.  When comparing our Arecibo sample to the \nan\ and \citet{papastergis2012} samples, we see the Arecibo sample follows roughly the same trend as the \nan\ data but with an approximately 0.5 dex offset toward lower \HI\ mass fractions.  From Paper I, the \nan\ regression fit is log(\MHI/\Mstar) = -0.59 log(\Mstar) + 5.05 and the Arecibo fit is log(\MHI/\Mstar) = -0.65 log(\Mstar) + 5.06.  Since the Arecibo sample is a small fraction of the total NIBLES sample, we only show this fit here to illustrate the differences in \HI\ mass fractions of the \nan\ undetected sample.  If this trend is representative of the rest of the NIBLES \nan\ undetected galaxies, we would expect the NRT regression fit to be offset toward lower mass fractions by about 0.17 dex.  Taking the entire stellar mass range into account and comparing mean \HI\ mass fraction values in 0.2 mag wide bins in log(\Mstar), we find that on average the Arecibo sample probes mass fractions that are 0.5 and 1 dex lower in log(\MHI) than the \nan\ and ALFALFA detections, respectively.

\section{Conclusions}

We obtained  about four times more sensitive follow-up \HI\ observations at Arecibo of 90 NIBLES galaxies that were either not detected or marginally detected at \nan.  We detected 80\% of these sources, which has enabled us to probe their underlying \HI\ distribution. The Arecibo detections have on average five times lower \HI\ masses than the \nan\ upper limits estimated in Paper I.  Contributing to this factor of five lower mass is not only the lower peak flux densities we are able to detect with Arecibo, but also the $\sim$37\% narrower line widths in our follow-up sample compared to the \nan\ detections of sources with the same optical luminosity.  This average difference in line width is primarily driven by the low luminosity (\Lr\ $<$ 10$^{8.5}$ \Lsun) sources which correspondingly show a higher central concentration of light.  This may be an indication that these relatively gas-poor galaxies have, on average, a more centrally confined \HI\ mass distribution compared to the \nan\ detected sample in the same luminosity range.

If we assume the $g$-$i$ color and \Lr\ distribution of Arecibo detection fractions are representative of the entire \nan\ undetected and marginally detected samples, we estimate $\sim$60\% (520) could be detected with the four times better sensitivity of our Arecibo observations.  This would put the over-all NIBLES detection rate at about 86\%.

Lastly, our Arecibo follow-up observations enabled us to sample our \nan\ undetected sample to \HI\ mass fractions 0.5 dex lower, on average, than our \nan\ detections.  Some of these galaxies with low \MHI/\Mstar\ fractions lie in virtually unexplored parameter space (e.g., around log(\Lr) = 11.5) and could potentially be used to shed further light on galaxy evolution processes studied by modelers, e.g., \cite{Kannappan2013}.

\begin{acknowledgements} 
The Arecibo Observatory is operated by SRI International under a cooperative agreement 
with the National Science Foundation (AST-1100968), and in alliance with Ana G. 
M\'endez-Universidad Metropolitana, and the Universities Space Research Association. 
The \nan\ Radio Telescope is operated as part of the Paris Observatory, in association with 
the Centre National de la Recherche Scientifique (CNRS) and partially supported by the 
R\'egion Centre in France. 
This research has made use of the HyperLeda database (http://leda.univ-lyon1.fr), the 
NASA/IPAC Extragalactic Database (NED) which is operated by the Jet Propulsion Laboratory, 
California Institute of Technology, under contract with the National Aeronautics and Space 
Administration, and the Sloan Digital Sky Survey which is managed by the Astrophysical Research 
Consortium for the Participating Institutions.
\end{acknowledgements}


\newpage
\clearpage
\onecolumn
\begin{landscape}
{\tiny
\tabcolsep=0.1cm
\begin{longtable}{lcclllllllllllllllllll}
\caption{\label{tab:AOdetNRTnd} Basic optical and \HI\ data -- Arecibo detections of \nan\ non-detections and marginals} \\

\hline\hline

Source & RA & Dec & Name & \Vopt\ & $g$-$z$  & \Mg\ & \Mstar\ & \sSFR\ & $rms$ & \VHI\ & \Wfifty\ & \Wtwenty\ & \FHI\  & $SNR$ & $S/N$ & \MHI\ & \MHI/\Mstar  \\ 

 & \multicolumn{2}{c}{(J2000.0)} & & & & & $[$log$]$ & $[$log$]$ & & & & & & & & $[$log$]$ & $[$log$]$ \\
 &	&     &    &   km/s &  mag &  mag & \Msun\ & yr$^{-1}$ & mJy & km/s & km/s & km/s & Jy km/s &  &  & \Msun\ & \\
\hline
\endfirsthead 
\caption{-- {\it continued}.} \\ 
\hline\hline 
Source & RA & Dec & Name & \Vopt\ & $g$-$z$ & \Mg\ & \Mstar\ & \sSFR\ & $rms$ & \VHI\ & \Wfifty\ & \Wtwenty\ & \FHI\ & $SNR$ & $S/N$ & \MHI\ & \MHI/\Mstar \\ 
& \multicolumn{2}{c}{(J2000.0)} & & & & & $[$log$]$ & $[$log$]$ & & & & & & & & $[$log$]$ & $[$log$]$ \\ & & & & km/s & mag & mag & \Msun\ & yr$^{-1}$ & mJy & km/s & km/s & km/s & Jy km/s & & & \Msun\ & \\ 
\hline
\endhead
\hline
\endfoot
\tabcolsep=0.07cm
$   0023^{}$  & 00 10 25.50 &  14 17 23.00  & \object{            KUG 0007+140} &  5452$\pm$4 &  0.93 & -18.81 &  9.50 &   -10.17 & 0.45 &  5456$\pm$3 & 189 &  208 &  0.60$\pm$0.06 &   8.4 &   15.4 &  8.93 & -0.57 \\
$  0037^{M}$  & 00 15 05.40 &  15 03 59.70  & \object{ 2MASX J00150533+1503597} &  5428$\pm$2 &  1.37 & -19.15 & 10.00 &   -10.55 & 0.53 &  5436$\pm$20 &  81 &  164 &  0.30$\pm$0.07 &   6.0 &   10.2 &  8.63 & -1.37 \\
$   0038^{}$  & 00 15 29.70 &  14 43 37.20  & \object{              ASK 147494} &  5432$\pm$3 &  0.39 & -16.55 &  8.21 &    -9.54 & 0.17 &  5445$\pm$3 &  65 &   76 &  0.04$\pm$0.01 &   4.9 &    4.7 &  7.77 & -0.44 \\
$   0297^{}$  & 02 00 03.80 &  14 07 12.90  & \object{               ASK 43581} &  4860$\pm$2 &  0.39 & -16.52 &  6.14 &    -8.70 & 0.57 &  4860$\pm$2 & 120 &  139 &  0.66$\pm$0.06 &  10.8 &   16.9 &  8.87 &  2.73 \\
$  0329^{M}$  & 02 50 52.00 &  00 00 12.10  & \object{              ASK 037212} &  1587$\pm$2 &  0.24 & -14.16 &  7.19 &    -9.59 & 0.17 &  1579$\pm$1 &  26 &   41 &  0.07$\pm$0.01 &  12.2 &   12.4 &  6.91 & -0.27 \\
$   0340^{}$  & 03 01 11.90 &  00 18 10.50  & \object{             PGC 1162351} &  2779$\pm$2 &  0.61 & -15.67 &  7.94 &    -9.52 & 0.54 &  2793$\pm$10 &  34 &   92 &  0.21$\pm$0.05 &   8.0 &   10.7 &  7.90 & -0.05 \\
$  0349^{M}$  & 03 07 15.65 &  00 43 52.00  & \object{               ASK 37834} &  3011$\pm$1 &  0.47 & -15.60 &  7.74 &    -9.43 & 0.58 &  3030$\pm$3 &  79 &  100 &  0.49$\pm$0.06 &  10.2 &   15.4 &  8.33 &  0.59 \\
$   0475^{}$  & 07 35 24.20 &  36 38 36.40  & \object{               ASK 44962} &  3902$\pm$1 &  0.42 & -16.86 &  8.23 &    -9.48 & 0.81 &  3887$\pm$8 &  86 &  141 &  0.79$\pm$0.09 &   9.9 &   17.1 &  8.76 &  0.53 \\
$   0562^{}$  & 07 59 16.20 &  23 50 58.10  & \object{              ASK 265000} &  2231$\pm$2 &  0.69 & -14.88 &  7.73 &    -9.96 & 0.40 &  2242$\pm$7 &  49 &   72 &  0.08$\pm$0.03 &   4.5 &    4.4 &  7.26 & -0.47 \\
$   0575^{}$  & 08 01 32.20 &  21 22 47.80  & \object{              ASK 363277} &  2104$\pm$3 &  0.56 & -15.63 &  7.86 &    -9.52 & 0.50 &  2103$\pm$3 &  47 &   78 &  0.46$\pm$0.04 &  13.6 &   21.8 &  7.99 &  0.13 \\
$   0602^{}$  & 08 15 38.03 &  21 15 37.20  & \object{              ASK 483519} &  3476$\pm$1 &  0.45 & -16.28 &  7.94 &    -9.31 & 0.51 &  3487$\pm$12 &  56 &  140 &  0.45$\pm$0.06 &   9.8 &   19.1 &  8.42 &  0.48 \\
$   0609^{}$  & 08 17 21.00 &  24 57 45.60  & \object{            KUG 0814+251} &  2074$\pm$4 &  0.61 & -15.55 &  7.91 &    -9.58 & 0.36 &  2062$\pm$7 &  88 &  131 &  0.29$\pm$0.04 &   8.4 &   14.0 &  7.77 & -0.13 \\
$   0645^{}$  & 08 23 41.00 &  20 41 48.50  & \object{              ASK 522849} &  2011$\pm$10 &  0.67 & -14.19 &  7.49 &    -9.75 & 0.45 &  2024$\pm$8 &  43 &   70 &  0.10$\pm$0.04 &   5.0 &    5.3 &  7.28 & -0.21 \\
$   0689^{}$  & 08 36 41.10 &  05 16 24.00  & \object{              ASK 259050} &  3127$\pm$1 &  0.20 & -15.72 &  7.54 &    -8.93 & 0.46 &  3126$\pm$5 &  55 &   90 &  0.28$\pm$0.04 &   9.5 &   13.3 &  8.12 &  0.58 \\
$  0708^{M}$  & 08 42 25.80 &  06 35 38.60  & \object{              ASK 259104} &  2260$\pm$1 &  0.33 & -15.48 &  7.60 &    -9.28 & 0.54 &  2254$\pm$4 &  50 &   85 &  0.47$\pm$0.05 &  12.4 &   19.8 &  8.06 &  0.46 \\
$   0719^{}$  & 08 44 42.70 &  29 32 43.00  & \object{              ASK 281794} &  2116$\pm$2 &  0.40 & -14.49 &  7.26 &    -9.29 & 0.50 &  2135$\pm$8 &  69 &  109 &  0.21$\pm$0.05 &   7.0 &    8.2 &  7.66 &  0.40 \\
$   0724^{}$  & 08 45 25.40 &  15 19 46.00  & \object{             PGC 4176159} &  1614$\pm$8 &  0.02 & -13.43 &  7.30 &   -10.72 & 0.36 &  1598$\pm$7 &  35 &   72 &  0.10$\pm$0.03 &   7.0 &    7.9 &  7.11 & -0.19 \\
$   0751^{}$  & 08 50 25.19 &  32 37 18.80  & \object{              ASK 266943} &  2221$\pm$1 &  0.56 & -14.82 &  7.21 &    -9.37 & 0.52 &  2219$\pm$5 &  44 &   63 &  0.10$\pm$0.04 &   4.9 &    4.9 &  7.39 &  0.18 \\
$  0801^{M}$  & 09 08 52.55 &  21 55 29.70  & \object{ 2MASX J09085253+2155292} &  3849$\pm$1 &  0.66 & -17.03 &  8.51 &    -9.63 & 0.59 &  3851$\pm$3 &  51 &   77 &  0.36$\pm$0.05 &  10.0 &   13.8 &  8.41 & -0.10 \\
$   0828^{}$  & 09 14 57.30 &  06 00 18.60  & \object{              ASK 261057} &  1370$\pm$29 &  0.64 & -14.37 &  ---- &     ---- & 0.32 &  1370$\pm$8 &  28 &   54 &  0.04$\pm$0.02 &   4.6 &    4.3 &  6.60 &  ---- \\
$  0850^{M}$  & 09 20 02.70 &  28 20 57.40  & \object{              ASK 488350} &  1915$\pm$7 &  0.52 & -14.23 &  7.35 &    -9.55 & 0.49 &  1920$\pm$3 &  63 &   83 &  0.29$\pm$0.04 &   8.6 &   11.9 &  7.70 &  0.35 \\
$   0904^{}$  & 09 40 45.80 &  32 28 20.80  & \object{            KUG 0937+327} &  1293$\pm$1 &  0.44 & -14.20 &  7.13 &    -9.32 & 0.51 &  1280$\pm$11 &  54 &  103 &  0.19$\pm$0.05 &   6.2 &    8.5 &  7.19 &  0.06 \\
$   0928^{}$  & 09 46 53.00 &  31 47 44.60  & \object{            KUG 0943+320} &  1425$\pm$3 &  0.64 & -14.40 &  7.47 &    -9.70 & 0.47 &  1424$\pm$5 &  59 &   92 &  0.27$\pm$0.04 &   8.4 &   12.3 &  7.43 & -0.04 \\
$  0938^{M}$  & 09 48 47.50 &  27 52 25.40  & \object{              ASK 491818} &  1316$\pm$1 &  0.51 & -14.56 &  7.30 &    -9.37 & 0.85 &  1318$\pm$10 &  34 &   73 &  0.20$\pm$0.07 &   5.5 &    6.6 &  7.22 & -0.07 \\
$   0949^{}$  & 09 52 35.10 &  08 11 56.60  & \object{              ASK 277328} &  2725$\pm$3 &  0.72 & -16.92 &  8.53 &    -9.75 & 0.56 &  2723$\pm$2 & 105 &  120 &  0.53$\pm$0.06 &   9.3 &   15.1 &  8.28 & -0.25 \\
$  0952^{M}$  & 09 54 30.00 &  32 03 42.00  & \object{              ASK 492645} &  1424$\pm$2 &  0.40 & -13.50 &  7.01 &    -9.47 & 0.66 &  1434$\pm$2 &  87 &  101 &  0.49$\pm$0.06 &   8.4 &   13.2 &  7.69 &  0.68 \\
$   0976^{}$  & 10 01 09.50 &  08 46 55.50  & \object{              ASK 277703} &  1265$\pm$2 &  0.63 & -13.67 &  7.14 &    -9.61 & 0.62 &  1260$\pm$2 &  28 &   39 &  0.13$\pm$0.04 &   7.0 &    6.5 &  7.00 & -0.14 \\
$  0987^{M}$  & 10 04 25.10 &  02 33 31.00  & \object{               ASK 68781} &  1125$\pm$2 & -0.38 & -11.91 &  7.12 &    -9.59 & 0.55 &  1131$\pm$2 &  32 &   53 &  0.33$\pm$0.04 &  12.5 &   17.2 &  7.30 &  0.19 \\
$  0991^{M}$  & 10 06 10.80 &  11 06 02.10  & \object{              ASK 295498} &  2499$\pm$1 &  0.48 & -16.95 &  6.10 &    -8.41 & 0.62 &  2509$\pm$3 & 101 &  123 &  0.58$\pm$0.07 &   9.1 &   15.2 &  8.24 &  2.15 \\
$   1017^{}$  & 10 16 59.00 &  03 42 35.40  & \object{               ASK 95821} &  1204$\pm$1 &  0.63 & -14.10 &  ---- &     ---- & 0.58 &  1212$\pm$3 &  29 &   46 &  0.17$\pm$0.04 &   8.2 &    8.7 &  7.07 &  ---- \\
$   1030^{}$  & 10 19 59.90 &  24 47 24.60  & \object{              ASK 596355} &  1258$\pm$1 &  0.42 & -14.06 &  7.08 &    -9.48 & 0.63 &  1256$\pm$4 &  35 &   56 &  0.17$\pm$0.04 &   7.4 &    7.3 &  7.10 &  0.02 \\
$   1069^{}$  & 10 35 11.10 &  25 27 04.00  & \object{             PGC 4243890} &  1298$\pm$1 &  0.24 & -13.32 &  6.62 &    -8.98 & 0.53 &  1297$\pm$4 &  34 &   56 &  0.16$\pm$0.04 &   7.4 &    8.4 &  7.11 &  0.49 \\
$   1095^{}$  & 10 44 20.70 &  14 05 04.00  & \object{                NGC 3357} &  9770$\pm$2 &  1.51 & -22.32 & 11.41 &   -12.52 & 0.77 &  9740$\pm$11 & 113 &  152 &  0.39$\pm$0.09 &   5.2 &    7.6 &  9.25 & -2.16 \\
$   1147^{}$  & 11 00 47.10 &  16 52 55.50  & \object{             PGC 4260762} &  1136$\pm$2 &  0.51 & -13.25 &  6.93 &    -9.56 & 1.06 &  1132$\pm$5 &  41 &   59 &  0.21$\pm$0.08 &   4.9 &    5.1 &  7.12 &  0.19 \\
$   1166^{}$  & 11 07 09.90 &  28 22 31.90  & \object{             PGC 1833985} &  1003$\pm$1 &  0.42 & -13.87 &  7.09 &    -9.49 & 0.48 &  1006$\pm$4 &  42 &   79 &  0.34$\pm$0.04 &  11.8 &   18.0 &  7.22 &  0.13 \\
$  1260^{D}$  & 11 25 04.00 &  28 09 35.60  & \object{             PGC 4546173} &  1593$\pm$1 & -0.01 & -14.24 &  6.76 &    -8.80 & 0.85 &  1580$\pm$9 &  38 &   65 &  0.16$\pm$0.07 &   4.2 &    4.8 &  7.27 &  0.51 \\
$   1388^{}$  & 11 47 06.90 &  03 06 23.00  & \object{               ASK 74477} &  1016$\pm$2 &  0.44 & -13.16 &  6.71 &    -9.29 & 0.19 &  1021$\pm$2 &  31 &   53 &  0.09$\pm$0.01 &  11.0 &   13.1 &  6.64 & -0.08 \\
$   1419^{}$  & 11 52 24.06 &  32 24 13.90  & \object{                NGC 3935} &  3066$\pm$2 &  1.37 & -19.60 & 10.22 &   -10.77 & 0.51 &  3075$\pm$19 &  78 &  152 &  0.27$\pm$0.06 &   5.7 &    9.9 &  8.09 & -2.12 \\
$   1424^{}$  & 11 53 00.30 &  16 02 29.40  & \object{              ASK 436517} &   905$\pm$29 &  0.63 & -12.08 &  ---- &     ---- & 0.58 &   931$\pm$4 &  27 &   42 &  0.09$\pm$0.04 &   4.6 &    5.0 &  6.59 &  ---- \\
$   1453^{}$  & 11 57 25.10 &  02 11 15.90  & \object{             PGC 1218832} &  1019$\pm$1 &  0.49 & -13.16 &  6.80 &    -9.49 & 0.62 &  1019$\pm$3 &  35 &   56 &  0.22$\pm$0.04 &   8.9 &    9.7 &  7.04 &  0.24 \\
$  1751^{M}$  & 12 54 12.60 &  00 48 09.10  & \object{            CGCG 015-035} &  1191$\pm$1 &  0.64 & -15.29 &  ---- &     ---- & 0.65 &  1197$\pm$2 &  61 &   81 &  0.62$\pm$0.06 &  12.6 &   20.0 &  7.63 &  ---- \\
$   1807^{}$  & 13 09 36.90 &  31 40 34.00  & \object{             PGC 1958740} &  2159$\pm$4 & -0.05 & -15.06 &  7.56 &    -9.91 & 0.25 &  2156$\pm$4 &  41 &   79 &  0.24$\pm$0.02 &  14.1 &   24.7 &  7.73 &  0.17 \\
$   1877^{}$  & 13 34 06.90 &  09 15 43.20  & \object{             PGC 4544337} &  1112$\pm$11 &  0.51 & -13.70 &  7.30 &   -10.09 & 0.26 &  1134$\pm$2 &  26 &   39 &  0.07$\pm$0.02 &   8.8 &    8.1 &  6.62 & -0.69 \\
$   1879^{}$  & 13 35 37.20 &  14 21 39.40  & \object{             PGC 1456087} &   993$\pm$21 &  0.78 & -13.81 &  ---- &     ---- & 0.45 &  1040$\pm$7 &  27 &   49 &  0.06$\pm$0.03 &   4.6 &    4.0 &  6.48 &  ---- \\
$  1987^{M}$  & 13 57 21.10 &  26 12 27.20  & \object{             PGC 1767195} &  2391$\pm$4 &  0.57 & -14.70 &  7.52 &    -9.55 & 0.32 &  2360$\pm$6 &  38 &   72 &  0.12$\pm$0.03 &   8.1 &   10.1 &  7.51 & -0.01 \\
$   1988^{}$  & 13 57 23.60 &  05 34 25.20  & \object{              ASK 179268} &   994$\pm$20 &  0.35 & -14.95 &  ---- &     ---- & 0.36 &  1008$\pm$1 &  42 &   55 &  0.17$\pm$0.03 &  10.3 &   11.9 &  6.93 &  ---- \\
$   1998^{}$  & 13 58 45.00 &  24 09 05.00  & \object{ 2MASX J13584501+2409048} &   970$\pm$2 &  0.76 & -13.86 &  7.34 &    -9.86 & 0.37 &   984$\pm$2 &  63 &   82 &  0.30$\pm$0.03 &  11.2 &   16.5 &  7.14 & -0.19 \\
$   2084^{}$  & 14 17 07.50 &  04 50 13.40  & \object{               ASK 99987} &  1643$\pm$1 &  0.37 & -14.11 &  7.09 &    -9.35 & 0.29 &  1658$\pm$3 &  45 &   73 &  0.18$\pm$0.02 &  10.7 &   14.6 &  7.37 &  0.28 \\
$   2094^{}$  & 14 19 29.08 &  35 34 01.10  & \object{              ASK 392553} &  3207$\pm$1 &  0.80 & -16.38 &  8.39 &    -9.80 & 0.30 &  3185$\pm$6 &  81 &  106 &  0.12$\pm$0.03 &   6.1 &    7.5 &  7.78 & -0.61 \\
$   2099^{}$  & 14 20 11.10 &  10 15 46.50  & \object{                NGC 5562} &  9142$\pm$3 &  1.61 & -21.84 & 11.40 &   -12.78 & 0.56 &  9187$\pm$4 & 140 &  149 &  0.18$\pm$0.07 &   3.4 &    4.4 &  8.87 & -2.53 \\
$   2140^{}$  & 14 28 08.70 &  01 49 25.60  & \object{              ASK 082514} &  1728$\pm$29 &  0.30 & -14.29 &  ---- &     ---- & 0.29 &  1728$\pm$3 &  25 &   43 &  0.08$\pm$0.02 &   8.6 &    8.9 &  7.06 &  ---- \\
$   2225^{}$  & 14 45 36.30 &  34 10 43.80  & \object{              ASK 394208} &  1642$\pm$6 &  0.40 & -14.51 &  7.15 &   -10.65 & 0.43 &  1630$\pm$3 &  35 &   57 &  0.21$\pm$0.03 &  10.8 &   13.2 &  7.42 &  0.27 \\
$  2236^{M}$  & 14 48 58.99 &  33 11 34.90  & \object{            CGCG 193-002} &  1700$\pm$2 &  0.70 & -16.54 &  8.33 &    -9.97 & 0.50 &  1704$\pm$2 &  50 &   77 &  0.64$\pm$0.04 &  15.2 &   28.9 &  7.95 & -0.38 \\
$  2301^{M}$  & 15 18 57.00 &  00 30 56.50  & \object{             PGC 1168006} &  2082$\pm$14 &  0.58 & -14.59 &  7.51 &    -9.63 & 0.36 &  2098$\pm$5 &  60 &   75 &  0.08$\pm$0.03 &   3.8 &    4.4 &  7.21 & -0.30 \\
$   2311^{}$  & 15 24 50.10 &  03 04 53.10  & \object{                SHOC 505} &  1753$\pm$1 &  0.38 & -14.87 &  7.56 &    -9.41 & 0.38 &  1745$\pm$5 &  80 &  125 &  0.40$\pm$0.04 &  11.2 &   19.3 &  7.77 &  0.22 \\
$   2316^{}$  & 15 27 44.50 &  09 41 56.80  & \object{              ASK 421256} &  1820$\pm$1 &  0.51 & -15.62 &  7.96 &    -9.67 & 0.41 &  1818$\pm$3 &  46 &   78 &  0.56$\pm$0.03 &  15.7 &   32.4 &  7.95 & -0.01 \\
$   2317^{}$  & 15 27 53.00 &  25 38 37.50  & \object{             PGC 1744110} &  1470$\pm$1 &  0.38 & -13.64 &  6.91 &    -9.26 & 0.31 &  1477$\pm$9 &  37 &   78 &  0.09$\pm$0.03 &   6.8 &    8.1 &  6.99 &  0.09 \\
$  2325^{M}$  & 15 32 49.90 &  36 12 13.50  & \object{              ASK 331970} &  1993$\pm$1 &  0.54 & -16.01 &  7.98 &    -9.56 & 0.50 &  2002$\pm$3 &  48 &   68 &  0.21$\pm$0.04 &   7.8 &    9.9 &  7.61 & -0.37 \\
$   2327^{}$  & 15 33 36.75 &  33 21 33.60  & \object{              ASK 313250} &  1999$\pm$2 &  0.36 & -14.56 &  7.40 &    -9.43 & 0.41 &  1994$\pm$8 &  52 &   93 &  0.19$\pm$0.04 &   7.3 &   10.5 &  7.56 &  0.16 \\
$   2349^{}$  & 16 00 13.80 &  17 50 53.80  & \object{             PGC 1543427} &  2053$\pm$1 &  0.46 & -15.64 &  7.71 &    -9.42 & 0.51 &  2057$\pm$3 &  79 &  106 &  0.59$\pm$0.05 &  11.9 &   21.1 &  8.08 &  0.37 \\
$   2373^{}$  & 16 35 20.70 &  17 45 55.10  & \object{                Mrk 0886} &  2740$\pm$1 &  0.63 & -17.82 &  8.94 &    -9.52 & 0.34 &  2754$\pm$5 &  86 &  129 &  0.39$\pm$0.04 &  11.3 &   19.9 &  8.15 & -0.79 \\
$   2402^{}$  & 21 13 07.70 &  01 13 47.20  & \object{ 2MASX J21130776+0113480} &  4157$\pm$2 &  0.75 & -18.01 &  8.97 &    -9.78 & 0.70 &  4165$\pm$10 & 146 &  187 &  0.49$\pm$0.09 &   5.8 &    9.3 &  8.61 & -0.36 \\
$  2407^{M}$  & 21 16 27.60 & -00 49 35.30  & \object{                NGC 7047} &  5783$\pm$3 &  1.48 & -20.60 & 10.71 &   -11.52 & 0.67 &  5796$\pm$6 & 422 &  458 &  2.11$\pm$0.14 &   8.4 &   24.7 &  9.53 & -1.18 \\
$  2410^{M}$  & 21 17 34.60 &  11 00 43.20  & \object{              ASK 138402} &  3271$\pm$2 &  0.47 & -15.37 &  7.50 &    -9.18 & 0.54 &  3266$\pm$6 &  59 &   99 &  0.36$\pm$0.05 &   9.4 &   14.3 &  8.27 &  0.77 \\
$   2416^{}$  & 21 28 29.10 &  10 04 52.60  & \object{ 2MASX J21282910+1004527} &  4929$\pm$2 &  1.02 & -17.63 &  9.10 &   -10.14 & 0.65 &  4933$\pm$5 & 178 &  192 &  0.32$\pm$0.09 &   3.9 &    5.9 &  8.57 & -0.53 \\
$   2423^{}$  & 21 42 22.80 &  12 29 53.80  & \object{               UGC 11794} &  5848$\pm$4 &  1.42 & -20.12 & 10.50 &   -11.28 & 0.46 &  5855$\pm$6 & 344 &  384 &  1.22$\pm$0.09 &   9.6 &   22.8 &  9.30 & -1.19 \\
$   2432^{}$  & 21 50 38.40 &  13 17 17.40  & \object{            CGCG 427-027} &  6045$\pm$2 &  1.50 & -20.81 & 10.88 &   -11.40 & 0.60 &  6060$\pm$6 & 208 &  231 &  0.41$\pm$0.09 &   5.6 &    7.8 &  8.87 & -2.02 \\
$   2433^{}$  & 21 52 22.50 & -01 10 15.70  & \object{               ASK 21400} &  4770$\pm$1 &  0.55 & -16.57 &  8.14 &    -9.48 & 1.31 &  4753$\pm$6 &  90 &  109 &  0.33$\pm$0.13 &   4.2 &    4.3 &  8.55 &  0.41 \\
$   2436^{}$  & 21 54 47.70 &  00 13 45.70  & \object{               ASK 21641} &  2984$\pm$1 &  0.48 & -15.76 &  7.88 &    -9.47 & 0.43 &  2981$\pm$4 &  73 &  107 &  0.47$\pm$0.04 &  12.4 &   20.9 &  8.30 &  0.42 \\
$   2452^{}$  & 22 26 19.50 &  12 15 02.00  & \object{              ASK 140946} &  3527$\pm$8 &  0.49 & -15.59 &  7.72 &    -9.51 & 0.32 &  3544$\pm$1 &  94 &  111 &  0.41$\pm$0.03 &  12.9 &   20.9 &  8.39 &  0.67 \\
$   2461^{}$  & 22 33 49.80 &  00 28 58.30  & \object{               ASK 23755} &  4642$\pm$1 &  0.47 & -16.75 &  8.22 &    -9.39 & 0.89 &  4606$\pm$22 &  59 &  161 &  0.52$\pm$0.11 &   6.7 &   12.2 &  8.72 &  0.51 \\
$   2534^{}$  & 23 25 30.20 &  14 06 20.50  & \object{              ASK 144855} &  3996$\pm$1 &  0.54 & -15.91 &  7.91 &    -9.49 & 0.29 &  4001$\pm$4 &  44 &   66 &  0.10$\pm$0.02 &   8.0 &    8.2 &  7.87 & -0.04 \\

\hline
\caption{\small Marginal \nan\ detections are flagged with an $M$ and sources affected by the baseline ripple with a $D$.  For a description of other flags, see Sect. \ref{results}.}
\end{longtable}
}

{\tiny
\tabcolsep=0.09cm

\begin{longtable}{lccllllllllllllllllll}
\caption{Basic optical and \HI\ data -- Arecibo marginal detections of \nan\ non-detections \label{tab:AOmarNRTnd}} \\
\hline\hline

Source & RA & Dec & Name & \Vopt\ & $g$-$z$  & \Mg\ & \Mstar\ & \sSFR\ & $rms$ & \VHI\ & \Wfifty\ &\Wtwenty\ & \FHI\  & $SNR$ & $S/N$ & \MHI\ & \MHI/\Mstar  \\ 

 & \multicolumn{2}{c}{(J2000.0)} & & & & & $[$log$]$ & $[$log$]$ & & & & & & & & $[$log$]$ & $[$log$]$ \\
 &	&     &    &   km/s &  mag &  mag & \Msun\ & yr$^{-1}$ & mJy & km/s & km/s & km/s & Jy km/s &  &  & \Msun\ & \\
\hline
$   0345^{}$  & 03 04 57.96 &  00 57 14.10  &  \object{                SHOC 150} &  3636$\pm$1 &  0.55 & -16.02 &  ---- &     ---- & 0.61 &  3634$\pm$3 &  25 &   34 &  0.05$\pm$0.03 &   3.6 &    2.8 &  7.52 &  ---- \\
$   1132^{}$  & 10 56 19.90 &  17 05 05.90  &  \object{             PGC 4257755} &   960$\pm$1 &  0.44 & -12.53 &  6.57 &    -9.47 & 0.65 &   966$\pm$12 &  17 &   37 &  0.03$\pm$0.04 &   2.6 &    1.9 &  6.14 & -0.43 \\
$   1163^{}$  & 11 06 32.10 &  11 23 07.50  &  \object{                NGC 3524} &  1357$\pm$1 &  1.32 & -18.37 &  9.71 &   -11.27 & 0.74 &  1341$\pm$8 & 415 &  431 &  0.23$\pm$0.14 &   2.7 &    2.5 &  7.30 & -2.40 \\
$   1989^{}$  & 13 57 29.52 &  09 57 03.20  &  \object{            CGCG 074-017} &  6969$\pm$2 &  1.30 & -19.82 & 10.29 &   -11.92 & 0.51 &  6922$\pm$10 & 317 &  344 &  0.21$\pm$0.09 &   3.7 &    3.6 &  8.68 & -1.61 \\
$   2162^{}$  & 14 31 53.00 &  03 22 48.30  &  \object{ 2MASX J23244466+0101490} &  1529$\pm$1 &  0.64 & -13.52 &  6.94 &    -9.37 & 0.47 &  1521$\pm$5 &  24 &   38 &  0.03$\pm$0.03 &   4.0 &    2.1 &  6.52 & -0.43 \\
 
\hline
\end{longtable}
}
\end{landscape}

{\tiny
\begin{longtable}{lcclllllllllll}
\caption{Basic optical and \HI\ data -- Arecibo non-detections \label{tab:AO_ND}} \\
\hline\hline

Source & RA & Dec & Name & \Vopt & $g$-$z$  & \Mg\ & \Mstar\ & \sSFR\ & $rms$ & $SNR$ & \MHI & \MHI/\Mstar \\ 
 & \multicolumn{2}{c}{(J2000.0)} & & & & & $[$log$]$ & $[$log$]$ & & & $[$log$]$ & $[$log$]$\\
 &	&     &    &   km/s &  mag &  mag & \Msun\ & yr$^{-1}$ & mJy & & \Msun\ & \\
\hline

$  0879^{K}$ & 09 34 02.80 &  10 06 31.30 & \object{                NGC 2914} & 3144$\pm$1 &  1.48 & -19.80 & 10.37 & -11.87 &  0.76 & 2.09 &  <8.89 & <-1.45 \\ 
$  0882^{M}$ & 09 35 05.80 &  09 38 57.10 & \object{ 2MASX J09350578+0938566} & 3408$\pm$2 &  1.27 & -18.12 &  9.61 & -11.36 &  0.61 & 2.80 &  <8.72 & <-0.58 \\ 
$   0906^{}$ & 09 41 16.60 &  10 38 49.10 & \object{                 IC 0552} & 5788$\pm$2 &  1.57 & -20.70 & 10.87 & -12.11 &  0.64 & 2.95 &  <9.31 & <-1.13 \\ 
$   1076^{}$ & 10 36 38.40 &  14 10 15.90 & \object{                NGC 3300} & 3017$\pm$1 &  1.42 & -20.24 & 10.53 & -12.29 &  0.49 & 2.18 &  <8.56 & <-1.45 \\ 
$   1146^{}$ & 11 00 35.40 &  12 09 41.60 & \object{                NGC 3491} & 6351$\pm$2 &  1.59 & -21.12 & 11.06 & -12.65 &  0.56 & 2.29 &  <9.11 & <-1.13 \\ 
$   1224^{}$ & 11 21 24.90 &  03 00 50.10 & \object{                NGC 3643} & 1742$\pm$2 &  1.19 & -17.98 &  9.53 & -11.41 &  0.88 & 2.05 &  <8.35 & <-1.05 \\ 
$   1849^{}$ & 13 24 10.00 &  13 58 35.50 & \object{                NGC 5129} & 6885$\pm$2 &  1.45 & -22.28 & 11.34 & -12.51 &  0.58 & 1.47 &  <9.20 & <-1.31 \\ 
$   1893^{}$ & 13 38 43.10 &  31 16 13.90 & \object{            CGCG 161-101} & 4699$\pm$2 &  1.53 & -19.85 & 10.52 & -12.39 &  0.60 & 2.62 &  <9.07 & <-1.17 \\ 
$   1951^{}$ & 13 52 26.70 &  14 05 28.60 & \object{                 IC 0948} & 6892$\pm$2 &  1.54 & -21.33 & 11.14 & -12.69 &  0.40 & 2.92 &  <9.01 & <-1.31 \\ 
$  2016^{K}$ & 14 02 48.60 &  09 20 28.90 & \object{                NGC 5423} & 5910$\pm$2 &  1.47 & -21.20 & 11.06 & -12.67 &  0.47 & 2.80 &  <9.09 & <-1.30 \\ 
$   2401^{}$ & 21 04 51.99 &  00 26 52.70 & \object{            CGCG 374-042} & 4129$\pm$2 &  1.36 & -18.88 &  9.95 & -12.02 &  0.63 & 2.35 &  <8.95 & <-0.34 \\ 
$   2406^{}$ & 21 16 24.80 &  10 16 24.10 & \object{            CGCG 426-029} & 5175$\pm$2 &  1.26 & -19.54 & 10.15 & -11.09 &  0.59 & 2.19 &  <9.16 & <-0.57 \\ 
$  2418^{M}$ & 21 31 37.60 &  11 49 53.90 & \object{            CGCG 426-062} & 8643$\pm$3 &  1.70 & -21.05 & 11.15 & -12.19 &  0.95 & 1.78 &  <9.87 & <-0.98 \\ 
$   2430^{}$ & 21 50 27.60 &  12 38 10.30 & \object{ 2MASX J21502753+1238103} & 6507$\pm$2 &  1.42 & -18.75 &  9.98 & -11.69 &  0.62 & 2.64 &  <9.33 & <-0.16 \\ 
$   2442^{}$ & 22 04 08.80 & -00 55 31.90 & \object{               ASK 22153} & 4825$\pm$16 &  0.82 & -16.68 &  8.66 & -10.35 &  0.15 & 3.14 &  <8.36 &  <0.86 \\

 \hline

\caption{\small Marginal and confused \nan\ detections are flagged with an $M$ and $K$ respectively.  The \MHI\ and \MHI/\Mstar columns list upper limit values.}
\end{longtable}
}

{\tiny
\tabcolsep=0.1cm
\begin{landscape}
\begin{longtable}{lcclllllllllllllllllll}

\caption{Basic optical and \HI\ data -- comparison of Arecibo and \nan\ detections
\label{tab:bothDETs}} \\

\hline\hline
Source & RA & Dex & Name & \Vopt\ & $g$-$z$  & \Mg\ & \Mstar\ & \sSFR\ & $rms$ & $rms$ & \VHI\ & \VHI\ & \Wfifty\ & \Wfifty\ &  \FHI\  & \FHI\ & $S/N$ & $S/N$ & \MHI\ & \MHI\ \\
 Telescope & & & & & &  & & & AO & NRT & AO & NRT & AO & NRT & AO & NRT & AO & NRT & AO & NRT\\
 & \multicolumn{2}{c}{(J2000.0)} & & & & & $[$log$]$ & $[$log$]$ & & & & & & & & & & & $[$log$]$ &  $[$log$]$\\ 
 &	&     &    &   km/s &  mag &  mag & \Msun\ & yr$^{-1}$ & mJy & mJy & km/s & km/s & km/s & km/s & Jy km/s & Jy km/s &  &  & \Msun\ & \Msun\ \\
\hline
\endfirsthead 
\caption{-- {\it continued}.} \\ 
\hline\hline 
Source & RA & DEC & Name & \Vopt\ & $g$-$z$ & \Mg\ & \Mstar\ & \sSFR\ & $rms$ & $rms$ & \VHI\ & \VHI\ & \Wfifty\ & \Wfifty\ & \FHI\ & \FHI\ & $S/N$ & $S/N$ & \MHI\ & \MHI\ \\ 
Telescope & & & & & & & & & AO & NRT & AO & NRT & AO & NRT & AO & NRT & AO & NRT & AO & NRT\\ 
& \multicolumn{2}{c}{(J2000.0)} & & & & & $[$log$]$ & $[$log$]$ & & & & & & & & & & & $[$log$]$ & $[$log$]$\\ 
& & & & km/s & mag & mag & \Msun\ & yr$^{-1}$ & mJy & mJy & km/s & km/s & km/s & km/s & Jy km/s & Jy km/s & & & \Msun\ & \Msun\ \\ 
\hline 
\endhead
\hline
\endfoot
\tabcolsep=0.09cm
$   0047^{}$ & 00 20 48.60 & 14 13 27.80 & \object{ 2MASX J00204857+1413283} & 5344$\pm$1 &  0.91 & -18.34 &  9.31 &      -9.51 &  0.61 &  1.69 &  5348 $\pm$7 &  5260 $\pm$7 & 107 &  55 &  0.65$\pm$0.07 &  0.3$\pm$0.1 &  16.51 &   4.03 &  8.95 &  8.63 \\ 
$   0337^{}$ & 02 59 14.50 & 00 33 59.70 & \object{               ASK 37367} & 2754$\pm$2 &  0.18 & -15.42 &  8.12 &       ---- &  0.59 &  2.21 &  2760 $\pm$1 &  2727 $\pm$11 &  53 &  56 &  1.17$\pm$0.05 &  0.5$\pm$0.2 &  44.25 &   4.76 &  8.63 &  8.24 \\ 
$   0576^{}$ & 08 01 58.90 & 21 22 19.10 & \object{              ASK 363290} & 2089$\pm$3 &  0.54 & -14.98 &  ---- &       ---- &  0.61 &  2.16 &  2087 $\pm$2 &  2058 $\pm$6 &  77 &  91 &  0.70$\pm$0.06 &  0.6$\pm$0.2 &  20.94 &   4.70 &  8.16 &  8.08 \\ 
$  0613^{K}$ & 08 17 53.93 & 24 41 12.00 & \object{              ASK 363797} & 2065$\pm$2 & -0.25 & -14.52 &  7.14 &      -8.98 &  0.53 &  1.54 &  2069 $\pm$2 &  2030 $\pm$2 &  49 & 134 &  1.83$\pm$0.05 & 10.1$\pm$0.2 &  79.47 &  93.39 &  8.58 &  9.31 \\ 
$ 0619^{C3}$ & 08 18 19.70 & 24 31 36.90 & \object{                 IC 2271} & 2213$\pm$2 &  0.45 & -17.07 &  8.29 &      -9.43 &  0.66 &  3.58 &  2227 $\pm$4 &  2208 $\pm$3 &  74 &  68 &  2.47$\pm$0.07 &  2.0$\pm$0.3 &  70.80 &  11.28 &  8.77 &  8.68 \\ 
$   0648^{}$ & 08 24 17.99 & 20 30 49.40 & \object{              ASK 522889} & 2166$\pm$3 &  0.69 & -16.44 &  7.69 &      -9.32 &  0.51 &  2.51 &  2170 $\pm$5 &  2146 $\pm$22 &  78 & 121 &  0.76$\pm$0.06 &  0.9$\pm$0.3 &  27.66 &   5.58 &  8.24 &  8.32 \\ 
$   0717^{}$ & 08 44 28.60 & 31 21 23.40 & \object{              ASK 282097} & 2200$\pm$5 &  0.50 & -15.63 &  7.84 &      -9.55 &  0.55 &  2.95 &  2199 $\pm$1 &  2175 $\pm$6 & 140 & 135 &  2.57$\pm$0.07 &  1.8$\pm$0.3 &  64.31 &   8.56 &  8.78 &  8.61 \\ 
$   0769^{}$ & 08 58 00.60 & 25 41 52.40 & \object{              ASK 486934} & 1897$\pm$0 &  0.34 & -15.13 &  7.43 &      -9.19 &  0.56 &  2.90 &  1888 $\pm$2 &  1878 $\pm$7 &  78 &  75 &  0.69$\pm$0.05 &  0.7$\pm$0.3 &  22.33 &   4.30 &  8.07 &  8.05 \\ 
$   0827^{}$ & 09 14 48.80 & 33 01 15.30 & \object{ 2MASX J09144880+3301148} & 1819$\pm$2 &  0.59 & -16.25 &  8.12 &      -9.49 &  0.57 &  2.95 &  1819 $\pm$9 &  1805 $\pm$8 &  46 &  76 &  0.40$\pm$0.06 &  0.4$\pm$0.3 &  16.66 &   2.75 &  7.80 &  7.83 \\ 
$   0854^{}$ & 09 21 14.99 & 09 43 52.20 & \object{              ASK 293255} & 1383$\pm$2 &  0.65 & -13.92 &  7.24 &      -9.62 &  0.57 &  2.45 &  1373 $\pm$4 &  1351 $\pm$4 &  46 &  48 &  0.16$\pm$0.04 &  0.3$\pm$0.2 &   6.61 &   2.55 &  7.15 &  7.37 \\ 
$0877^{CKR}$ & 09 33 46.10 & 10 09 09.00 & \object{                NGC 2911} & 3231$\pm$2 &  1.73 & -20.46 & 11.04 &     -13.08 &  0.79 &  1.60 &  3205 $\pm$22 &  3106 $\pm$67 & 422 & 314 &  4.72$\pm$0.19 &  2.3$\pm$0.4 &  47.42 &  13.63 &  9.37 &  9.05 \\ 
$   1000^{}$ & 10 12 52.80 & 22 43 19.20 & \object{            CGCG 123-024} & 1281$\pm$3 &  0.80 & -16.05 &  8.16 &      -9.73 &  0.64 &  3.08 &  1293 $\pm$2 &  1277 $\pm$6 & 109 & 104 &  1.27$\pm$0.07 &  0.6$\pm$0.3 &  30.83 &   3.10 &  8.01 &  7.67 \\ 
$  1025^{K}$ & 10 18 10.30 & 07 08 34.70 & \object{              ASK 230324} & 3779$\pm$2 &  0.68 & -17.39 &  8.71 &      -9.80 &  0.81 &  2.68 &  3777 $\pm$8 &  3666 $\pm$17 & 121 & 184 &  1.55$\pm$0.11 &  4.8$\pm$0.4 &  28.04 &  21.65 &  9.03 &  9.50 \\ 
$   1047^{}$ & 10 25 46.40 & 05 39 06.70 & \object{            CGCG 037-033} & 1154$\pm$2 &  0.35 & -14.88 &  7.38 &     -10.54 &  0.55 &  2.24 &  1155 $\pm$1 &  1146 $\pm$8 &  97 & 102 &  2.89$\pm$0.06 &  2.4$\pm$0.3 &  86.72 &  17.31 &  8.27 &  8.18 \\ 
$   1126^{}$ & 10 54 21.90 & 27 14 22.10 & \object{                NGC 3451} & 1356$\pm$1 &  1.04 & -18.19 &  ---- &       ---- &  0.63 &  2.55 &  1335 $\pm$1 &  1326 $\pm$3 & 238 & 235 &  7.27$\pm$0.10 &  6.1$\pm$0.4 & 122.40 &  25.90 &  8.80 &  8.72 \\ 
$  1156^{K}$ & 11 03 59.80 & 27 56 16.10 & \object{             PGC 4571781} & 1424$\pm$2 &  0.69 & -13.45 &  6.99 &      -9.70 &  0.65 &  2.28 &  1413 $\pm$17 &  1366 $\pm$2 & 103 & 191 &  0.64$\pm$0.09 &  4.4$\pm$0.3 &  15.64 &  22.93 &  7.79 &  8.60 \\ 
$   1161^{}$ & 11 05 32.50 & 17 38 22.50 & \object{ 2MASX J11053253+1738228} & 1039$\pm$2 &  0.75 & -16.24 &  8.30 &     -10.00 &  0.77 &  1.56 &  1042 $\pm$1 &  1030 $\pm$5 &  97 &  95 &  2.06$\pm$0.08 &  1.6$\pm$0.2 &  44.29 &  16.92 &  8.03 &  7.90 \\ 
$   1188^{}$ & 11 12 39.80 & 09 03 21.00 & \object{                 IC 0676} & 1412$\pm$3 &  1.37 & -18.26 &  9.69 &     -10.30 &  1.27 &  1.85 &  1433 $\pm$7 &  1430 $\pm$26 & 188 & 145 &  1.66$\pm$0.18 &  1.2$\pm$0.3 &  15.57 &   8.81 &  8.21 &  8.07 \\ 
$  1238^{K}$ & 11 22 50.70 & 12 20 41.50 & \object{                 IC 2781} & 1547$\pm$1 &  0.44 & -15.17 &  7.45 &      -9.35 &  0.61 &  2.39 &  1546 $\pm$2 &  1558 $\pm$13 &  82 & 106 &  1.26$\pm$0.06 &  1.4$\pm$0.3 &  36.95 &   9.50 &  8.16 &  8.22 \\ 
$   1242^{}$ & 11 23 07.00 & 30 28 44.10 & \object{            KUG 1120+307} & 1610$\pm$1 &  0.61 & -16.73 &  8.27 &      -9.48 &  0.57 &  3.27 &  1617 $\pm$2 &  1611 $\pm$9 & 104 &  95 &  2.86$\pm$0.06 &  2.6$\pm$0.4 &  79.80 &  13.26 &  8.56 &  8.51 \\ 
$   1291^{}$ & 11 31 08.90 & 13 34 13.40 & \object{              ASK 433348} & 1019$\pm$2 &  0.40 & -13.31 &  6.77 &      -9.29 &  0.74 &  1.09 &  1021 $\pm$2 &  1010 $\pm$5 &  38 &  50 &  0.49$\pm$0.05 &  0.3$\pm$0.1 &  17.30 &   6.11 &  7.39 &  7.15 \\ 
$   1296^{}$ & 11 31 44.60 & 34 20 00.20 & \object{               UGC 06512} & 1871$\pm$3 &  0.58 & -17.87 &  8.69 &      -9.56 &  0.75 &  1.81 &  1866 $\pm$1 &  1853 $\pm$0 & 155 & 154 &  9.32$\pm$0.09 &  8.1$\pm$0.2 & 161.74 &  60.00 &  9.19 &  9.13 \\ 
$  1349^{K}$ & 11 41 29.80 & 32 20 59.50 & \object{                Mrk 0746} & 1804$\pm$1 &  0.31 & -16.69 &  7.94 &      -9.16 &  0.60 &  3.05 &  1842 $\pm$5 &  1806 $\pm$4 & 107 & 129 &  3.60$\pm$0.07 &  8.4$\pm$0.4 &  94.92 &  39.99 &  8.77 &  9.12 \\ 
$   1426^{}$ & 11 54 01.60 & 16 43 24.00 & \object{              PGC 166116} &  983$\pm$1 & -0.43 & -13.59 &  7.04 &      -9.45 &  0.57 &  2.76 &   978 $\pm$1 &   970 $\pm$4 &  32 &  35 &  0.89$\pm$0.04 &  0.8$\pm$0.2 &  44.28 &   8.55 &  7.61 &  7.59 \\ 
$   1427^{}$ & 11 54 04.40 & 30 06 34.60 & \object{             PGC 4301309} &  980$\pm$9 &  0.18 & -13.61 &  6.76 &      -9.41 &  0.54 &  3.56 &   970 $\pm$1 &   968 $\pm$2 &  68 &  69 &  2.93$\pm$0.05 &  2.4$\pm$0.3 & 108.38 &  13.33 &  8.12 &  8.03 \\ 
$   1501^{}$ & 12 04 04.50 & 28 58 58.20 & \object{            KUG 1201+292} &  920$\pm$1 &  0.78 & -15.58 &  ---- &       ---- &  0.46 &  1.17 &   912 $\pm$2 &   908 $\pm$14 &  69 &  62 &  0.69$\pm$0.04 &  0.5$\pm$0.1 &  29.26 &   8.16 &  7.44 &  7.26 \\ 
$   1528^{}$ & 12 08 24.20 & 03 00 47.90 & \object{              ASK 075814} &  880$\pm$2 &  0.64 & -12.39 &  6.58 &      -9.56 &  0.78 &  1.75 &   884 $\pm$4 &   859 $\pm$6 &  31 &  33 &  0.10$\pm$0.05 &  0.2$\pm$0.1 &   3.80 &   2.70 &  6.58 &  6.77 \\ 
$   1572^{}$ & 12 13 48.30 & 12 41 25.90 & \object{                 IC 3052} &  812$\pm$14 &  0.38 & -13.60 &  7.03 &     -10.16 &  0.64 &  3.59 &   847 $\pm$1 &   834 $\pm$9 &  61 &  85 &  0.67$\pm$0.05 &  0.7$\pm$0.3 &  21.98 &   3.27 &  7.36 &  7.34 \\ 
$   1625^{}$ & 12 21 07.00 & 00 27 40.80 & \object{                 ASK 570} & 1894$\pm$1 &  0.27 & -14.81 &  7.20 &      -8.96 &  0.72 &  1.60 &  1890 $\pm$4 &  1875 $\pm$7 &  39 &  49 &  0.43$\pm$0.06 &  0.4$\pm$0.1 &  15.50 &   5.54 &  7.87 &  7.81 \\ 
$  1631^{R}$ & 12 22 03.90 & 09 02 05.60 & \object{                NGC 4307} &  962$\pm$14 &  1.63 & -17.74 &  ---- &       ---- &  0.53 &  1.79 &  1053 $\pm$2 &  1055 $\pm$20 & 310 & 331 &  1.16$\pm$0.09 &  1.2$\pm$0.3 &  20.56 &   6.09 &  7.79 &  7.81 \\ 
$   1633^{}$ & 12 22 08.20 & 15 47 56.40 & \object{                VCC 0530} & 1301$\pm$10 & -0.08 & -14.91 &  ---- &       ---- &  0.65 &  1.51 &  1296 $\pm$1 &  1288 $\pm$5 &  36 &  42 &  0.60$\pm$0.05 &  0.6$\pm$0.1 &  24.92 &  10.64 &  7.69 &  7.71 \\ 
$   1790^{}$ & 13 05 18.80 & 36 06 10.40 & \object{                 IC 4171} & 1201$\pm$3 &  0.48 & -15.62 &  7.71 &      -9.93 &  0.77 &  2.94 &  1206 $\pm$1 &  1197 $\pm$2 & 115 & 111 &  4.99$\pm$0.08 &  3.8$\pm$0.3 &  98.61 &  20.36 &  8.54 &  8.42 \\ 
$   1897^{}$ & 13 39 22.34 & 31 14 57.60 & \object{              ASK 526959} &  703$\pm$2 &  0.28 & -13.50 &  6.72 &      -9.14 &  0.48 &  2.45 &   702 $\pm$2 &   694 $\pm$3 &  41 &  51 &  2.17$\pm$0.04 &  1.8$\pm$0.2 & 115.15 &  16.92 &  7.71 &  7.62 \\ 
$   2068^{}$ & 14 12 58.55 & 09 55 15.90 & \object{            CGCG 074-140} & 7054$\pm$2 &  1.26 & -20.11 & 10.32 &      -9.94 &  0.49 &  1.79 &  7079 $\pm$2 &  6910 $\pm$11 & 290 & 291 &  1.80$\pm$0.08 &  1.6$\pm$0.3 &  34.94 &   8.60 &  9.64 &  9.59 \\ 
$   2176^{}$ & 14 35 33.30 & 12 54 29.60 & \object{               UGC 09389} & 1835$\pm$2 &  0.67 & -17.92 &  8.86 &     -10.16 &  0.40 &  1.34 &  1826 $\pm$1 &  1811 $\pm$0 & 229 & 230 & 16.32$\pm$0.06 & 14.4$\pm$0.2 & 438.20 & 117.39 &  9.42 &  9.36 \\ 
$   2178^{}$ & 14 35 50.10 & 02 36 19.70 & \object{               ASK 83319} & 1569$\pm$9 &  0.47 & -15.23 &  6.18 &       ---- &  0.50 &  3.48 &  1557 $\pm$1 &  1550 $\pm$5 &  63 &  56 &  2.19$\pm$0.04 &  1.8$\pm$0.3 &  89.09 &  11.68 &  8.41 &  8.33 \\ 
$   2224^{}$ & 14 45 33.60 & 31 54 56.10 & \object{              ASK 470206} & 1208$\pm$14 & -0.61 & -13.60 &  7.12 &     -10.73 &  0.62 &  1.91 &  1212 $\pm$1 &  1203 $\pm$5 &  74 &  75 &  0.94$\pm$0.06 &  0.9$\pm$0.2 &  28.84 &   8.56 &  7.82 &  7.78 \\ 
$   2247^{}$ & 14 52 43.50 & 11 40 19.90 & \object{              ASK 417578} & 1803$\pm$4 &  0.63 & -15.32 &  7.75 &      -9.44 &  0.54 &  3.40 &  1802 $\pm$1 &  1784 $\pm$5 & 123 & 109 &  2.10$\pm$0.06 &  1.5$\pm$0.4 &  57.57 &   7.09 &  8.52 &  8.37 \\ 
$   2341^{}$ & 15 55 22.40 & 02 55 15.10 & \object{              ASK 104991} & 1998$\pm$7 &  0.30 & -14.44 &  7.23 &      -9.54 &  0.50 &  2.90 &  2002 $\pm$1 &  1993 $\pm$8 &  65 &  72 &  0.67$\pm$0.04 &  0.6$\pm$0.3 &  26.84 &   3.91 &  8.11 &  8.05 \\ 
$   2414^{}$ & 21 23 18.40 & 01 15 18.10 & \object{ 2MASX J21231841+0115175} & 5458$\pm$2 &  1.51 & -18.72 & 10.08 &     -10.43 &  0.90 &  1.51 &  5467 $\pm$9 &  5397 $\pm$21 & 236 & 227 &  1.32$\pm$0.14 &  1.1$\pm$0.2 &  15.46 &   7.68 &  9.28 &  9.19 \\ 
$  2417^{K}$ & 21 30 25.88 & -00 28 27.70& \object{ 2MASX J21302589-0028272} & 5964$\pm$2 &  1.48 & -18.56 &  9.91 &     -10.10 &  0.67 &  1.74 &  6008 $\pm$5 &  5905 $\pm$21 & 267 & 269 &  2.00$\pm$0.11 &  2.6$\pm$0.3 &  29.54 &  15.12 &  9.54 &  9.66 \\
$   2419^{}$ & 21 35 03.60 & 10 57 36.20 & \object{              ASK 138954} & 3490$\pm$7 &  0.63 & -16.60 &  8.06 &      -9.60 &  0.30 &  2.30 &  3502 $\pm$1 &  3465 $\pm$35 & 128 &  96 &  0.77$\pm$0.03 &  0.6$\pm$0.3 &  37.06 &   4.53 &  8.66 &  8.57 \\ 
$   2421^{}$ & 21 41 11.20 & 12 43 15.80 & \object{              ASK 139485} & 6206$\pm$3 &  0.52 & -17.81 &  8.81 &      -9.72 &  0.49 &  1.50 &  6207 $\pm$1 &  6080 $\pm$13 & 161 & 139 &  0.64$\pm$0.06 &  0.7$\pm$0.2 &  16.60 &   6.52 &  9.08 &  9.12 \\ 
$   2428^{}$ & 21 45 22.60 & 12 16 06.30 & \object{              ASK 139520} & 5776$\pm$4 &  0.36 & -16.80 &  8.21 &      -9.41 &  0.56 &  1.73 &  5799 $\pm$3 &  5689 $\pm$8 & 118 & 148 &  0.49$\pm$0.06 &  0.4$\pm$0.2 &  12.87 &   3.46 &  8.90 &  8.86 \\ 
$2434^{CKD}$ & 21 52 37.90 & 12 32 09.00 & \object{            CGCG 427-032} & 8724$\pm$3 &  1.83 & -20.88 & 11.16 &     -11.90 &  1.49 &  2.58 &  8540 $\pm$13 &  8469 $\pm$7 &  93 & 139 &  1.01$\pm$0.18 &  0.5$\pm$0.3 &  11.11 &   2.76 &  9.55 &  9.28 \\ 
$   2435^{}$ & 21 54 17.99 & 00 56 31.50 & \object{ 2MASX J21541799+0056318} & 2976$\pm$1 &  0.42 & -17.39 &  8.36 &      -9.18 &  0.61 &  2.73 &  2991 $\pm$8 &  2943 $\pm$10 & 138 & 186 &  1.58$\pm$0.09 &  1.2$\pm$0.4 &  35.68 &   5.36 &  8.83 &  8.71 \\ 
$   2439^{}$ & 22 00 44.10 & 12 18 03.00 & \object{              ASK 140028} & 8800$\pm$1 &  0.27 & -18.13 &  6.89 &      -9.45 &  0.47 &  1.70 &  8785 $\pm$6 &  8544 $\pm$5 & 128 & 142 &  1.07$\pm$0.06 &  0.9$\pm$0.2 &  32.10 &   6.90 &  9.60 &  9.51 \\ 
$   2441^{}$ & 22 03 15.99 & 00 34 15.90 & \object{                NGC 7189} & 9020$\pm$1 &  1.45 & -21.51 & 11.10 &     -10.63 &  0.63 &  3.00 &  9054 $\pm$3 &  8774 $\pm$38 & 391 & 393 &  2.58$\pm$0.12 &  3.2$\pm$0.6 &  33.29 &   8.79 & 10.01 & 10.10 \\ 
$ 2449^{CK}$ & 22 21 33.70 & 12 31 22.30 & \object{              ASK 140519} & 7771$\pm$1 &  0.89 & -18.50 &  ---- &       ---- &  0.46 &  1.79 &  7769 $\pm$5 &  7578 $\pm$13 & 405 & 364 &  1.86$\pm$0.09 &  1.8$\pm$0.3 &  32.30 &   8.63 &  9.73 &  9.72 \\ 
$   2450^{}$ & 22 22 08.80 & 12 04 24.10 & \object{              ASK 141077} & 5125$\pm$4 &  0.71 & -16.44 &  8.68 &      -9.46 &  0.53 &  1.53 &  5108 $\pm$1 &  5026 $\pm$11 & 155 & 149 &  1.20$\pm$0.07 &  0.8$\pm$0.2 &  29.02 &   6.81 &  9.18 &  8.99 \\ 
$   2464^{}$ & 22 39 31.31 & -00 40 36.10& \object{               ASK 23608} & 7728$\pm$3 &  0.58 & -17.19 &  ---- &       ---- &  0.82 &  1.83 &  7719 $\pm$2 &  7507 $\pm$11 & 158 & 146 &  0.99$\pm$0.10 &  0.6$\pm$0.2 &  15.44 &   4.33 &  9.45 &  9.23 \\
$  2469^{K}$ & 22 40 57.58 & -01 15 08.70& \object{             PGC 1123197} & 4747$\pm$3 &  1.48 & -18.99 & 10.10 &     -11.15 &  0.77 &  1.33 &  4760 $\pm$5 &  4734 $\pm$9 &  78 & 211 &  0.36$\pm$0.07 &  0.7$\pm$0.2 &   8.46 &   5.58 &  8.59 &  8.86 \\

 \hline
\caption{\small AO = Arecibo, NRT = \nan.  Flags used in column Source: $C$ for clear confusion with another galaxy within the Arecibo telescope beam, $C3$ for another galaxy in the beam that is unlikely to cause confusion and $K$ for sources confused within the \nan\ telescope beam.}

\end{longtable}
\end{landscape}

}

{\tiny
\tabcolsep=0.09cm
\begin{longtable}{lcclllllllllllllllll}
\caption{Basic optical and \HI\ data -- Arecibo detected galaxies not included in the final \nan\ sample \label{tab:additional}} \\

\hline\hline
Source & RA & Dec & Name & \Vopt\ & $g$-$z$  & \Mg\ & \Mstar\ & \sSFR\ & $rms$ & \VHI\ & \Wfifty\ & \Wtwenty\ & \FHI\  & $SNR$ & $S/N$ & \MHI\ & \MHI/\Mstar  \\ 

 & \multicolumn{2}{c}{(J2000.0)} & & & & & $[$log$]$ & $[$log$]$ & & & & & & & & $[$log$]$ & $[$log$]$ \\
 &	&     &    &   km/s &  mag &  mag & \Msun\ & yr$^{-1}$ & mJy & km/s & km/s & km/s & Jy km/s &  &  & \Msun\ & \\
\hline
\tabcolsep=0.09cm
$   2601^{}$  & 01 21 02.95 &  00 51 05.30  &  \object{   ASK 032535} &  2438$\pm$1 &  0.43 & -15.61 &  7.71 &    -9.40 & 0.88 &  2433$\pm$6 &  25 &   37 &  0.06$\pm$0.05 &   2.6 &    2.0 &  7.20 & -0.51 \\
$   2602^{}$  & 08 17 07.90 &  24 33 45.70  &  \object{   ASK 363798} &  2127$\pm$2 &  0.42 & -15.27 &  6.30 &    -9.65 & 0.52 &  2126$\pm$1 &  63 &   79 &  1.39$\pm$0.04 &  18.1 &   54.6 &  8.48 &  2.18 \\
$   2603^{}$  & 08 18 50.20 &  22 06 55.30  &  \object{ CGCG 119-044} &  3484$\pm$2 &  0.58 & -17.88 &  8.74 &    -9.53 & 0.63 &  3494$\pm$1 & 124 &  138 &  1.88$\pm$0.07 &  16.2 &   43.1 &  9.04 &  0.31 \\
$   2604^{}$  & 08 46 47.29 &  13 42 24.40  &  \object{ CGCG 061-011} &  2141$\pm$3 &  0.77 & -16.59 &  8.47 &    -9.89 & 0.55 &  2142$\pm$1 &  66 &   89 &  1.46$\pm$0.05 &  18.1 &   52.8 &  8.51 &  0.04 \\
$   2605^{}$  & 08 52 31.44 &  00 51 12.80  &  \object{   ASK 058363} &  3252$\pm$1 &  0.62 & -15.41 &  7.81 &    -9.70 & 0.48 &  3259$\pm$3 &  80 &   95 &  0.13$\pm$0.04 &   6.2 &    4.8 &  7.81 &  0.01 \\
$   2606^{}$  & 11 37 48.50 &  22 41 28.60  &  \object{     NGC 3772} &  3551$\pm$2 &  1.42 & -19.75 & 10.35 &   -11.73 & 0.39 &  3423$\pm$7 & 112 &  137 &  0.16$\pm$0.04 &   4.8 &    6.2 &  7.95 & -2.40 \\
$   2607^{}$  & 11 40 13.90 &  24 41 49.40  &  \object{     NGC 3798} &  3567$\pm$2 &  1.47 & -20.39 & 10.61 &   -10.70 & 0.74 &  3552$\pm$4 & 389 &  415 &  1.97$\pm$0.14 &   9.8 &   21.9 &  9.08 & -1.53 \\
$   2608^{}$  & 14 18 53.47 &  09 17 28.70  &  \object{   ASK 456832} &  1201$\pm$1 & -1.17 & -13.37 &  6.94 &    -9.40 & 0.28 &  1208$\pm$2 &  49 &   71 &  0.28$\pm$0.02 &  14.0 &   23.5 &  7.30 &  0.36 \\
$   2609^{}$  & 14 45 20.20 &  34 19 48.10  &  \object{   ASK 394205} &  1666$\pm$5 &  0.38 & -14.83 &  7.35 &    -9.25 & 0.61 &  1668$\pm$2 &  45 &   75 &  0.84$\pm$0.05 &  16.5 &   33.6 &  8.05 &  0.71 \\
$   2610^{}$  & 21 30 59.86 & -00 00 02.10  &  \object{ CGCG 375-048} &  9034$\pm$3 &  0.95 & -20.72 & 10.22 &   -10.55 & 0.97 &  9052$\pm$2 & 228 &  252 &  3.36$\pm$0.15 &  13.7 &   36.6 & 10.12 & -0.09 \\

\hline
\caption{\small Sources which were excluded from the original data release in Paper I due to observational problems.}
\end{longtable}

}


\bibstyle{aa}
\bibliographystyle{aa}

\bibliography{NIBLES_bib}

\newpage
\clearpage 

\clearpage

\begin{appendix}

\section{Baseline derippling}
\label{app:derip}

\begin{figure}[!h] 
\centering
\includegraphics[width=9cm]{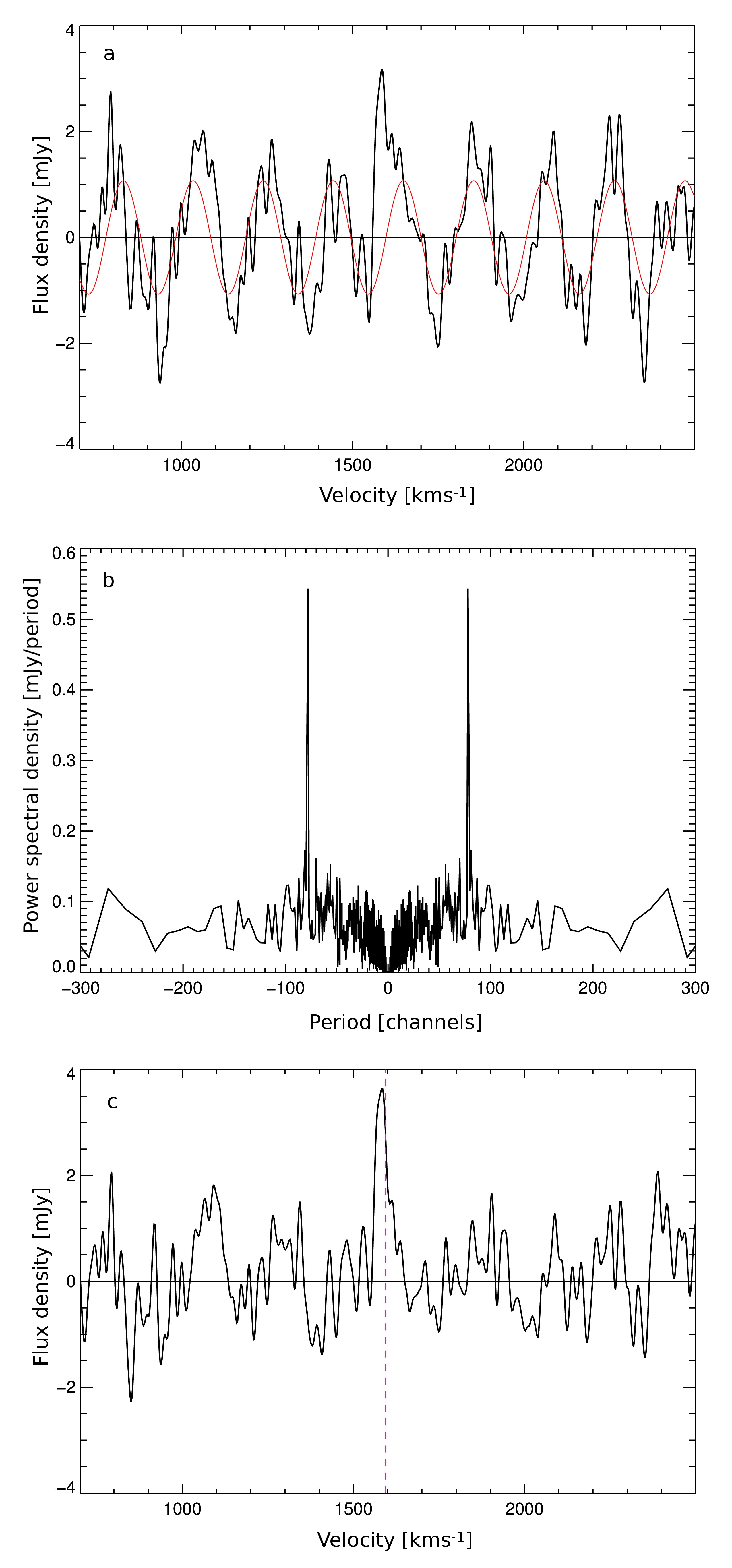}
\caption{\small Illustration of the removal of a baseline ripple, in source 1260. {\bf a.}  Original observed \HI\ line spectrum, flux density (mJy) as a function of heliocentric velocity with the standing wave over-layed in red.  {\bf b.} Spectral power density (mJy/period) vs. period for the Fourier transform where the spikes in power density corresponding to 78 channels due to the baseline ripple are clearly much higher than the rest of the spectrum.  {\bf c.} The reverse Fourier transform of {\bf b} with the 78 channel period peaks removed.  The standing wave has clearly been eliminated from the spectrum and the source signal is now easily identified.  The vertical magenta dashed line indicates the SDSS optical velocity.  This illustration represents one of two cases in our sample where this procedure was used.  Source 1260 contained the stronger of the two standing waves.}

\label{fft_multiple}
\end{figure}

Two of our sources, 1260 and 2434 (i.e., \object{PGC 4546173} and \object{CGCG 427-032}), suffered from a baseline ripple with a wavelength of approximately 210 \kms\ due to reflected radiation in the telescope structure \cite[see][]{briggs1997, wilson09}.  At Arecibo, this effect is caused by the formation of a standing wave between the primary mirror and receiver cabin.  It can be caused by a number of phenomena but is typically the result of a strong continuum source or broadband terrestrial RFI.  The ripple shows up in the spectrum as a result of a slight phase variation between the ON and OFF scans.  In Fourier space, this manifests itself as a narrow spike corresponding to a period across 78 correlator channels or a wavelength of $\sim$300m (frequency of $\sim$1 MHz).  This is wavelength is exactly what one would expect to see in a standing wave since they form in multiples of half-wavelength distances between two reflecting surfaces.  The distance between the primary mirror and the receiver cabin is 150m.

To illustrate this phenomenon, in Fig. \ref{fft_multiple}a we show the spectrum of source 1260 as it originally appeared with the standing wave easily apparent, which we have over-layed in red for reference.  The Fourier transform (Fig. \ref{fft_multiple}b) shows the effect of the ripple as two clearly identifiable spikes. After removing the offending period and doing an inverse Fourier transform (Fig. \ref{fft_multiple}c), the source signal is much more easily identified, resulting in a detection.

\newpage

\section{Arecibo spectra}
\label{app:spec}

\begin{figure}[ht]
\centering
\includegraphics[page=1, width=14cm]{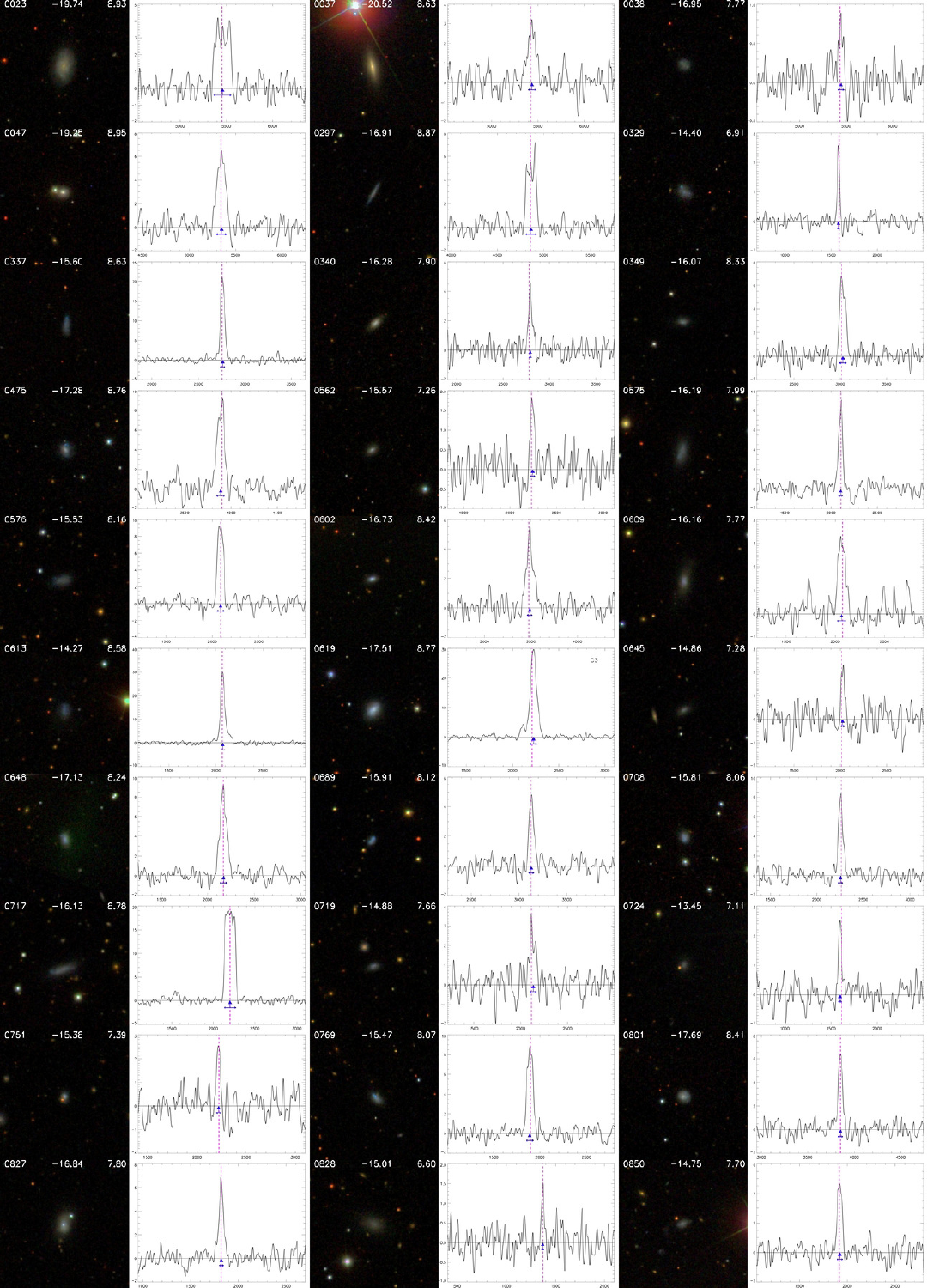}
\captcont{\small {\bf a.} Color ($g$,$r$ and $i$ band composite) images from the SDSS alongside 21-cm \HI\ line spectra of galaxies clearly detected at Arecibo.  The size of each image is 2\arcmin $\times$ 2\arcmin with the NIBLES source number indicated in the upper left corner, absolute z-band magnitude, \Mz, in the top center and log(\MHI) in \Msun in the top right corner of each image.  The vertical axis in the spectra is flux density in mJy, the horizontal axis is heliocentric radial velocity ($cz$) in \kms.  The SDSS recession velocity is denoted by a vertical dashed magenta line, the mean \HI\ velocity by the blue triangle and the \Wfifty\ line width by the horizontal blue arrow bar.  Confused galaxies are denoted by their confusion code from Sect. \ref{results} in the upper right portion of the spectrum.  Velocity resolution is 18.7 \kms.}
\label{AO_dets:1}
\end{figure}

\begin{figure}
\centering
\includegraphics[width=17cm, page=2]{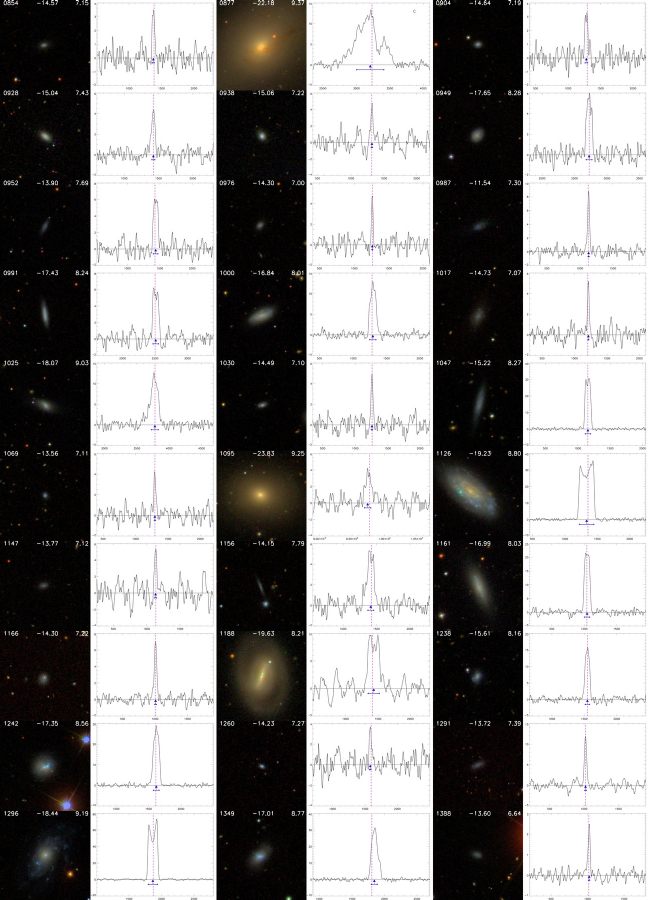}

\captcont*{\small {\bf b.} Color images from the SDSS alongside the 21-cm \HI\ line spectra of galaxies clearly detected at Arecibo (cont.).}

\label{AO_dets:2}
\end{figure}

\begin{figure}
\centering
\includegraphics[width=17cm, page=3]{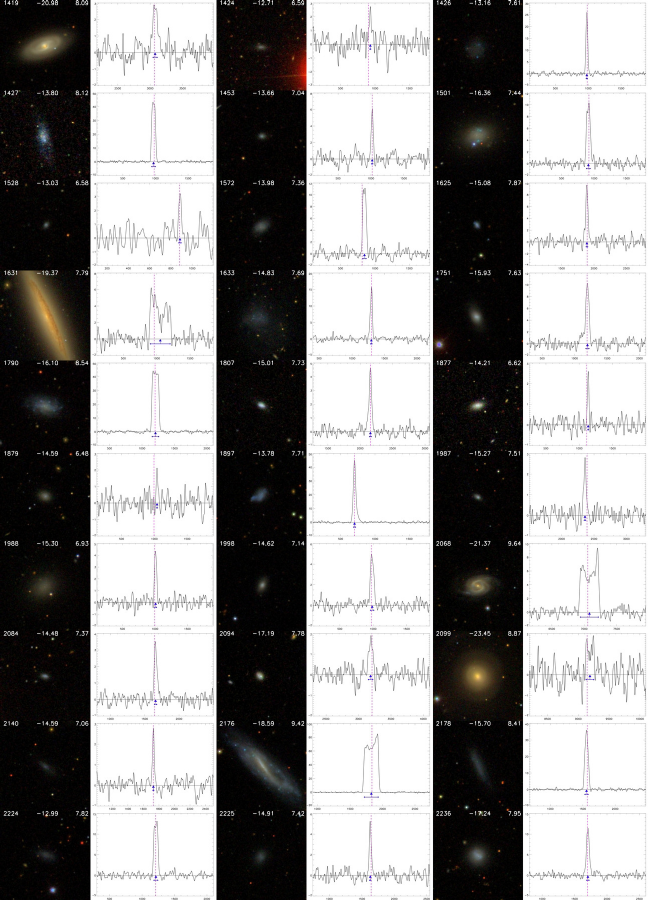}

\captcont*{\small {\bf c.} Color images from the SDSS alongside the 21-cm \HI\ line spectra of galaxies clearly detected at Arecibo (cont.).}

\label{AO_dets:3}
\end{figure}

\begin{figure}
\centering
\includegraphics[width=17cm, page=4]{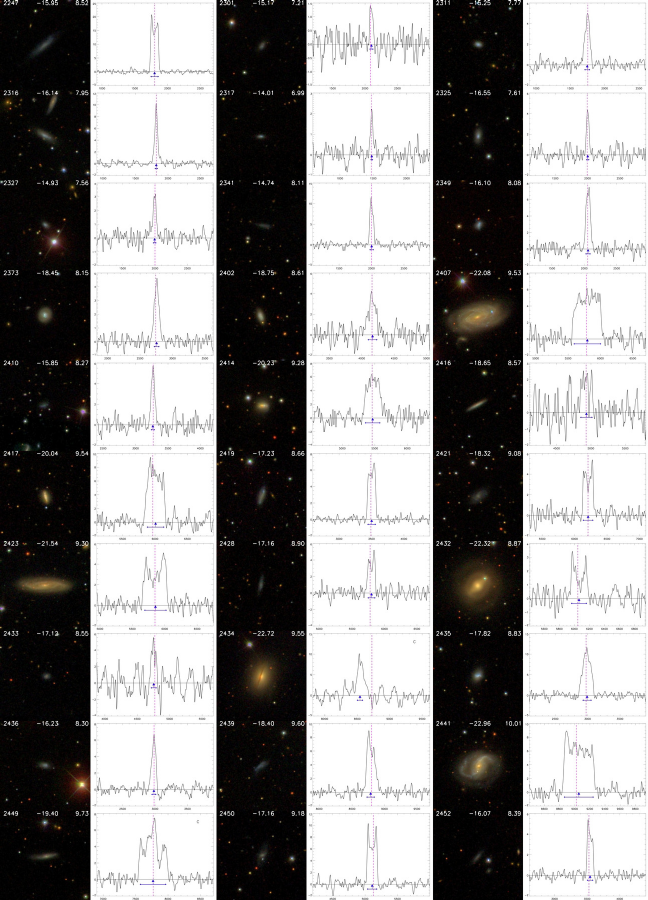}

\captcont*{\small {\bf d.} Color images from the SDSS alongside the 21-cm \HI\ line spectra of galaxies clearly detected at Arecibo (cont.).}

\label{AO_dets:4}
\end{figure}

\begin{figure}
\centering
\includegraphics[width=17cm, page=5]{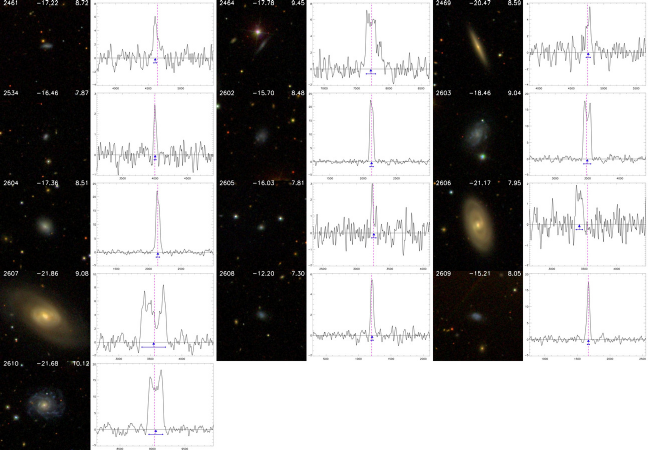}

\captcont*{\small {\bf e.} Color images from the SDSS alongside the 21-cm \HI\ line spectra of galaxies clearly detected at Arecibo (cont.).}

\label{AO_dets:5}
\end{figure}

\begin{figure}
\centering
\addtocounter{figure}{1}
\includegraphics[width=17cm]{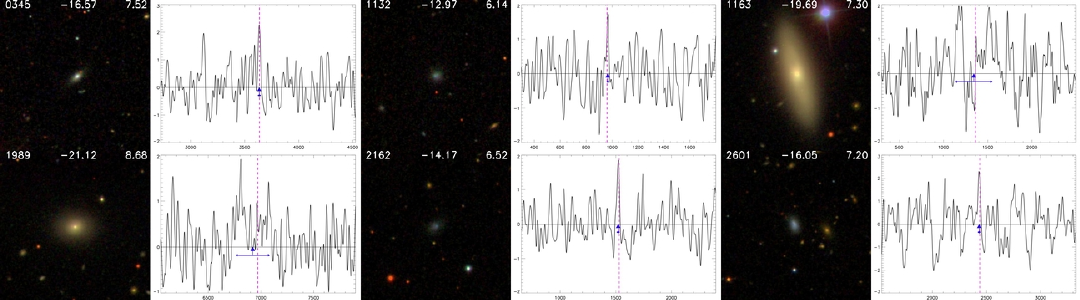}
\caption{\small Color images from the SDSS alongside the 21-cm \HI\ line spectra of galaxies marginally detected at Arecibo. See Figure \ref{AO_dets:1} for further details.}

\label{AO_mars}
\end{figure}

\begin{figure}
\centering
\includegraphics[width=17cm]{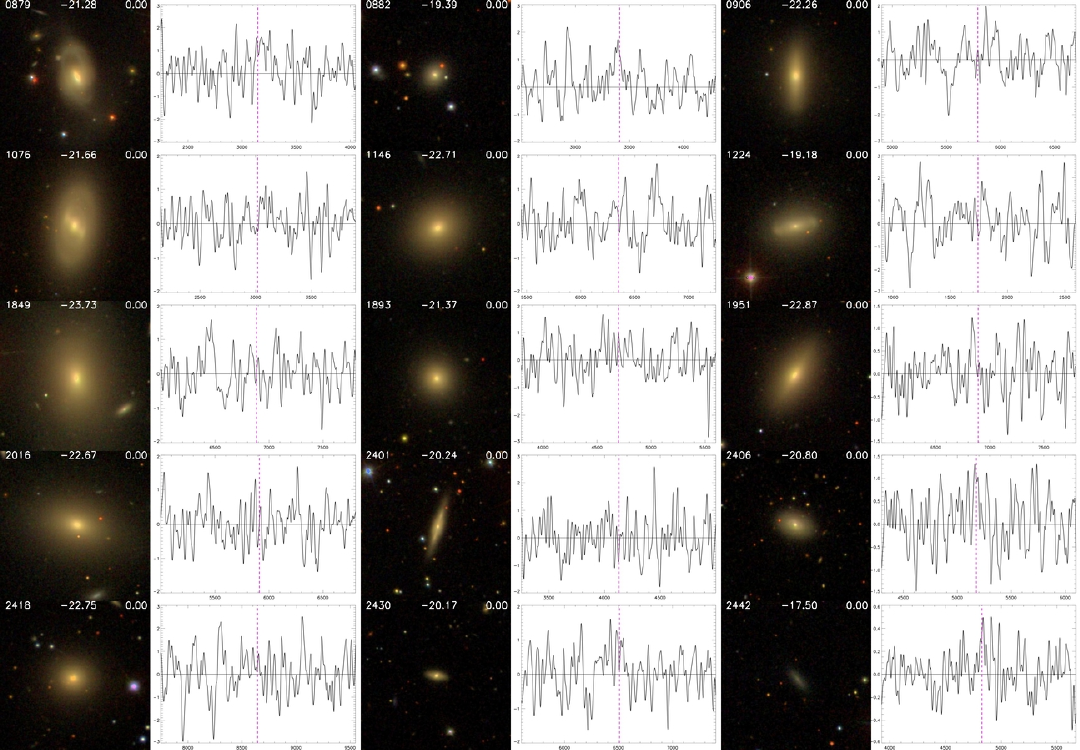}
\caption{\small Color images from the SDSS alongside the 21-cm \HI\ line spectra of galaxies undetected at Arecibo. See Figure \ref{AO_dets:1} for further details.}

\label{AO_ND}
\end{figure}

\end{appendix}

\end{document}